\documentclass[preprint2,tighten]{aastex6}
\pdfoutput=1 %for arXiv submission
\usepackage{amsmath,amstext}
\usepackage{apjfonts}
\usepackage[T1]{fontenc}
\usepackage[figure,figure*]{hypcap} %Figure refs go figures
\usepackage{enumitem}
\include{hyperlink-year-only-natbib-patch}

\def\sek~{\S\,}

\shorttitle{A MegaCam Survey of Halo Satellites}
\shortauthors{Mu\~noz et al.}

\begin{document}

%\shorttitle{A MegaCam Survey of Outer Halo Satellites}
%\shortauthors{Mu\~noz et al.}

\title{A MegaCam Survey of Outer Halo Satellites. I. 
Description of the Survey\altaffilmark{1,2}}

\author{Ricardo\ R.\ Mu\~noz\altaffilmark{3,4}}
\author{Patrick\ C\^ot\'e\altaffilmark{5}}
\author{Felipe\ A.\ Santana\altaffilmark{3}}
\author{Marla\ Geha\altaffilmark{4}}
\author{Joshua\ D.\ Simon\altaffilmark{6}}
\author{Grecco\ A.\ Oyarz\'un\altaffilmark{3}}
\author{Peter\ B. Stetson\altaffilmark{5}}
\and
\author{S.\ G.\ Djorgovski\altaffilmark{7}}

\altaffiltext{1}{Based on observations obtained at the Canada-France-Hawaii Telescope (CFHT) which is operated by the National Research Council of 
Canada, the Institut National des Sciences de l'Univers of the Centre National de la Recherche Scientifique of France,  and the University of Hawaii.}
\altaffiltext{2}{This paper includes data gathered with the 6.5 meter Magellan Telescopes located at Las Campanas Observatory, Chile.}
\altaffiltext{3}{Departamento de Astronom\'ia, Universidad de Chile, Camino del Observatorio 1515, Las Condes, Santiago, Chile (rmunoz@das.uchile.cl)}
\altaffiltext{4}{Astronomy Department, Yale University, New Haven, CT 06520, USA}
\altaffiltext{5}{National Research Council of Canada, Herzberg Astronomy and Astrophysics, 5071 W. Saanich Road, Victoria, BC V9E 2E7, Canada}
\altaffiltext{6}{Observatories of the Carnegie Institution for Science, 813 Santa Barbara St., Pasadena, CA 91101, USA}
\altaffiltext{7}{Astronomy Department, California Institute of Technology, Pasadena, CA, 91125, USA}

\begin{abstract}
We describe a deep, systematic imaging study of satellites in the outer halo of the  Milky Way. Our sample consists of 58 
stellar overdensities --- i.e., substructures classified as either globular clusters, classical dwarf galaxies, or ultra-faint dwarf galaxies ---
that are located at  Galactocentric distances of $R_{\rm GC} \ge 25$~kpc (outer halo) and out to $\sim400$\,kpc. 
This includes 44 objects for which we have 
acquired deep, wide-field, $g$- and $r$-band imaging with the MegaCam mosaic cameras on the 
3.6m Canada-France-Hawaii Telescope and the 6.5m Magellan-Clay telescope. 
These data are supplemented by archival imaging, or published $gr$ photometry, for an additional 14 objects, most of which were 
discovered recently in the Dark Energy Survey (DES). We describe the scientific motivation for our 
survey, including sample selection, observing strategy, data reduction pipeline, calibration procedures, and the depth and precision of 
the photometry. The typical 5$\sigma$ point-source limiting magnitudes for our MegaCam imaging --- which collectively covers an
area of $\approx 52$~deg$^2$ --- are $g_{\rm lim} \simeq 25.6$ and $r_{\rm lim} \simeq 25.3$ AB mag. These limits are comparable 
to those from the coadded DES images and are roughly a half-magnitude deeper than will be reached in a single visit with LSST. Our 
photometric catalog thus provides the deepest and most uniform photometric database  of Milky Way satellites available for the 
foreseeable future. In other papers in this series, we have used these data to explore the blue straggler populations in these objects, their 
density distributions, star formation histories, scaling relations and possible foreground structures.
\end{abstract}


\section{Introduction}
\label{sec:intro}

In the currently favored $\Lambda$CDM paradigm, the fossil record of galaxy formation is imprinted
in the stellar halos that surround massive galaxies like the Milky Way. The number of substructures, or satellites, embedded in the halo is an important prediction of cosmological models --- and one that depends sensitively on 
a number of parameters: the small-scale power spectrum, the assumed properties of the dark matter particles and a variety of baryonic processes that can modify the shape of the satellite luminosity function.
In principle, a comparison between the predicted and  observed number of satellites can 
provide a straightforward test of cosmological models \citep{moore99a,klypin99a}. In practice, though, there 
are significant technical challenges and serious selection effects involved in finding and characterizing 
Galactic satellites. 

Such satellites have historically been separated into two types of objects thought to 
have had dissimilar origins: globular clusters 
and dwarf galaxies. In recent years, it has become fashionable to further subdivide the latter category 
on the basis of luminosity or surface brightness: i.e., ``classical" versus ``ultra-faint" dwarf galaxies,
although these are not physically distinct classes.
Beginning with the SDSS \citep{york00a} --- which provided uniform  $ugriz$ photometry covering a significant 
fraction of the sky ---  a large number of faint dwarf galaxies and globular clusters 
have been discovered. 
The number of known satellites 
has increased steadily for more than two centuries.
Historically, sharp 
increases in the number of satellites have followed soon after the deployment of powerful new
survey facilities (e.g., the telescopes of Herschel, the Oschin Schmidt, the SDSS, Pan-STARRS,
and, most recently, the Dark Energy Camera on the Blanco 4m telescope). 

As of early 2017, the 
Milky Way is known to contain at least 77 satellites located beyond a Galactocentric radius of 
25~kpc.
It is virtually certain that this number will continue to rise as observing facilities improve. In
the next  decade, the highly anticipated Large Synoptic  Survey Telescope (LSST) should be especially
important in improving our census of satellites, at least in the southern hemisphere.
 
In the meantime, it is desirable to characterize the properties of the known satellites in a careful and systematic
way. Such efforts are presently hampered 
by the lack of uniform, high-quality imaging for these objects, which are scattered over the entire sky and span
wide ranges in apparent size, luminosity and surface brightness. Existing catalogs containing photometric and
structural parameters for the Milky Way satellites have tended to focus on either globular clusters and dwarf 
galaxies, despite the fact that the distinction between these stellar systems has become increasingly
blurred over the years (i.e., at the lowest surface brightnesses, separating these populations is often impossible
without spectroscopic information). Moreover, previous compilations continue to rely on shallow and heterogenous 
data (with some of the early ones even photographic), some of it dating back to the 1960s 
(see, e.g., \citealt{djorgovski93,pryor93,trager95a,irwin95a,harris96a,mateo98a,mclaughlin05,mcconnachie12}
and references therein).

Deep, homogeneous, digital photometry for a nearly complete sample of halo satellites would allow a
fresh look into the nature of these satellites. Issues of particular interest include possible connections between
the various sub-populations (globular clusters vs. classical dwarfs vs. ultra-faint dwarfs), including their density 
distributions \citep{hubble30,sersic68,elson87,king62a,king66,plummer11a,wilson75}, 
stellar populations, dark matter content, and evidence for tidal interactions with the Milky Way.
Here we introduce a deep, wide-field imaging survey of satellites belonging to the outer 
Galactic halo. The survey, which makes no {\it a priori} selection on the basis of object morphology 
or classification, is based on homogeneous $g$- and $r$-band imaging for 44 objects obtained 
with mosaic CCD cameras on the  3.6m Canada-Hawaii-France Telescope (CFHT) and the 6.5m 
Magellan/Clay telescope.  Our imaging is supplemented with photometry assembled from the 
literature --- or derived from our reduction and analysis of archival imaging --- for an additional 14 objects. 
In 2015, when the analysis of our secondary targets was completed, the survey provided
uniform photometry for a nearly complete ($\gtrsim$95\%) sample of the 60 satellites located more than 25~kpc 
from the Galactic center. Since that time, a number of additional satellites have been detected, 
primarily by the Dark Energy Survey (DES), so our sample now represents roughly three quarters of the 
77 cataloged members of the outer halo.

Our survey is the first to combine depth, wide areal coverage, 
multi-band imaging, and a high level of completeness for outer halo satellites. In this paper, we 
describe the survey strategy, sample selection, data reduction and calibration.
Previous papers in this series have examined the internal dynamics of the unusual stellar 
system {\tt Palomar~13} \citep{bradford11}, reported the discovery of {\tt Munoz~1}, an extremely 
low-luminosity star cluster in the field of the {\tt Ursa Minor} dwarf galaxy \citep{munoz12b}, studied 
the properties of blue straggler stars in remote satellites \citep{santana13}, examined possible 
foreground populations in the direction of {\tt NGC2419} and {\tt Koposov~2} \citep{carballobello15},
and characterized the Sagittarius tidal stream in the vicinity of the globular cluster {\tt Whiting~1} \citep{carballobello17}.
A companion paper \citep{munoz18a} presents structural parameters homogeneously derived and
future papers will explore the density distributions of our sample objects and compare their photometric 
and structural parameters to those of other stellar systems.

This paper is organized as follows. In \S\ref{sec:samp}, we discuss the sample selection for our survey. 
In \S\ref{sec:obs_primary}, we describe the observing strategy and data reduction procedures for our primary 
sample of 44 objects observed with the CFHT or Clay telescopes. In \S\ref{sec:obs_secondary}, we discuss the 
assembly of published or archival data for our secondary sample of 14 recently discovered objects 
that were not included in the primary sample.  \S\ref{sec:checks} discusses some consistency
checks on our photometry, while \S\ref{sec:results} presents some illustrative
results from our program.  We summarize and conclude in \S\ref{sec:summary}.

\section{Selection of Targets}
\label{sec:samp}

The goal of our survey is a homogeneous study of the photometric and structural parameters for a 
large, unbiased sample of satellites residing in the outer halo of the Milky Way. This is a significant
undertaking as it requires deep,
uniform, wide-field, and multi-filter imaging for dozens of objects that span a wide range in luminosity, surface
brightness and distance, and are scattered across the northern and southern skies. Initially,
our survey was designed to rely entirely on CFHT and Magellan imaging for 41 Galactic
satellites --- a complete list of Galactic satellites as of 2009 (excluding Sagittarius and the Magellanic Clouds).
Henceforth, we shall refer to this sample --- and three more objects that were discovered after the start of the 
survey --- as our ``primary sample" (see \S\ref{sec:primary}). In \S\ref{sec:secondary}, we discuss 
a ``secondary sample" of 14 satellites discovered in 2013, 2014 or 2015 that relies on published or archival data.  

\begin{figure}[t]
  \includegraphics[width=0.5\textwidth]{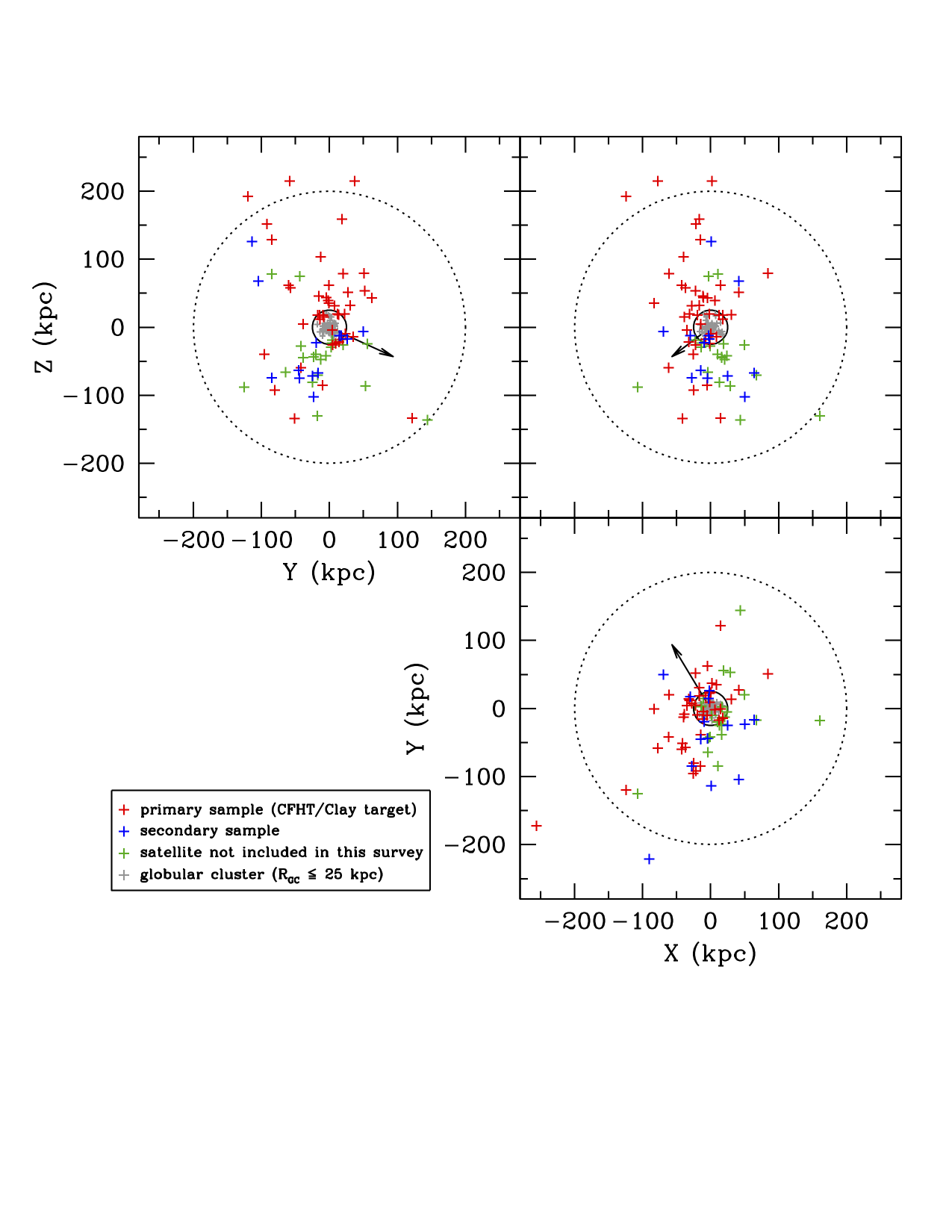}
\caption{Distribution of Milky Way satellites in a Cartesian coordinate system centered on the Galactic center. The 44 objects in
our primary sample (i.e., having CFHT or Clay imaging) are indicated by the red crosses. Blue crosses indicate objects in our 
expanded sample: i.e., recently discovered satellites whose structural parameters are examined using archival or literature data. 
Gray crosses show Galactic globular clusters while green crosses indicate known satellites as of 2017 that are absent from
our study. The inner circle at 25~kpc indicates the Galactocentric radius that defines the boundary of study of ``outer halo" 
substructures. The dashed circle at 200~kpc shows the virial radius of the Milky Way (e.g., Dehnen et~al. 2006).}
\label{fig:xyz}
\end{figure}

\subsection{Primary Sample}
\label{sec:primary}

In an important departure from most previous studies, we make no {\it a priori} selection on satellite
morphology or classification. This seems prudent since, as discussed in \S\ref{sec:intro}, the once-clear distinction
between globular clusters, classical dwarf galaxies and ultra-faint dwarf galaxies has become
increasingly blurred in recent years. Rather, our sample is defined purely on the basis of location in 
the Galactic halo, with all overdensities located beyond a Galactocentric distance of $R_{\rm GC} = 25$~kpc 
considered appropriate targets. Although somewhat arbitrary, this choice for the inner boundary of
the ``outer halo" seems reasonable given that inside this radius, halo stars tend to have higher 
metallicities and different kinematics than their more distance counterparts 
(see, e.g.,~\citealt{carollo07a,carollo10a} and references therein).
While we impose no firm cutoff on {\it outer} radius, we do confine ourselves to satellites that are
believed to be members of the Milky Way satellite system (e.g., \citealt{mateo98a,mcconnachie12}). Our
most distant satellite, {\tt Leo~T}, lies at $D_{\odot} \approx 417$~kpc --- a point at which the distinction 
between membership in the Galactic and Local Group satellite subsystems becomes unclear. Our
next most distant objects --- {\tt Leo~I} and {\tt Leo~II} --- lie at distances of 254 and 233~kpc,
respectively. Thus, our survey focuses on the population of satellites between 
$R_{\rm GC} = 25$~kpc and the virial radius of the Milky Way, which various estimates place between 
200 and 300~kpc (e.g., \citealt{klypin02,dehnen06,xue08}).

 \begin{figure}[t]
  \includegraphics[width=0.49\textwidth]{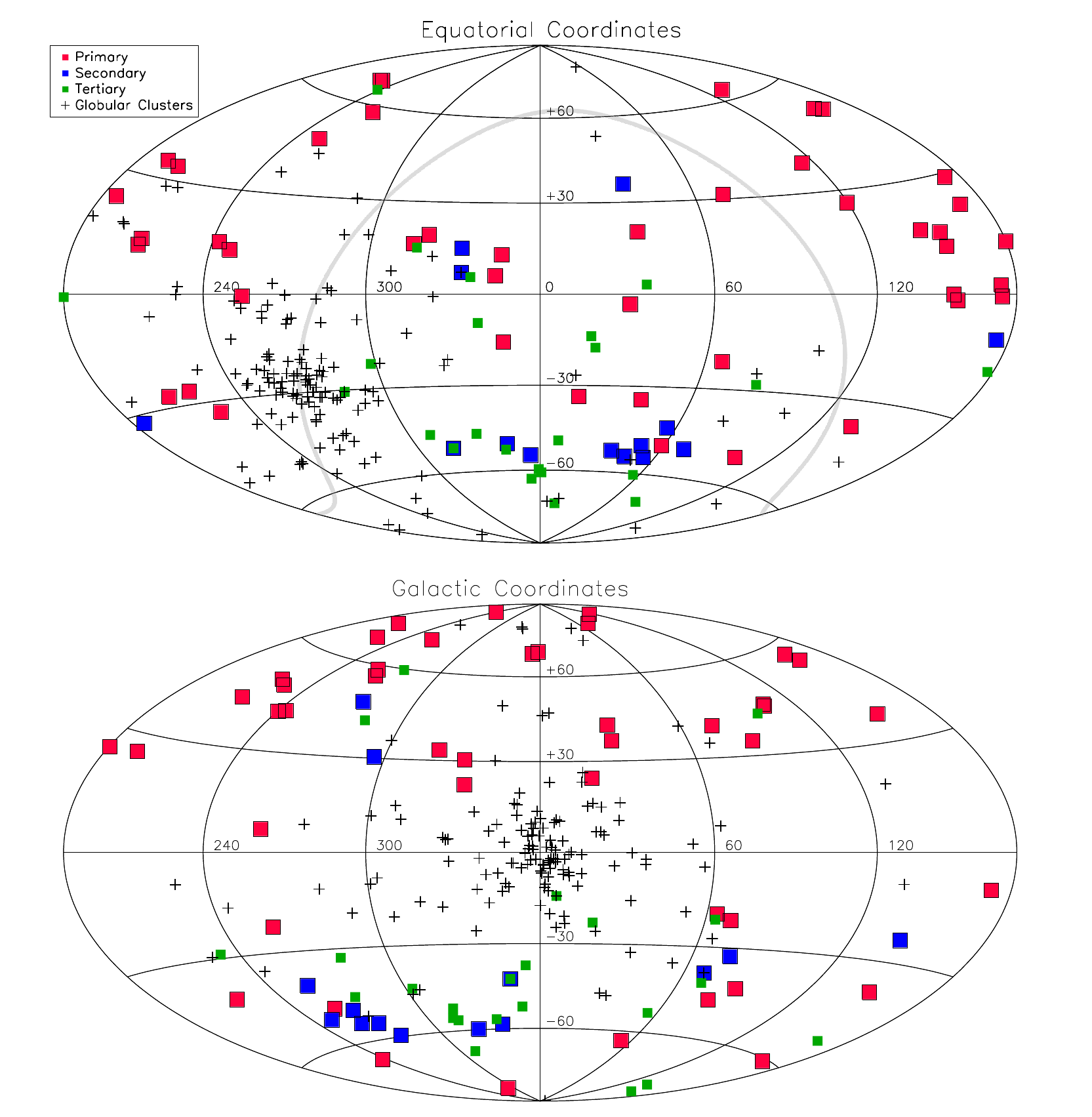}
\caption{{\it upper panel:} Aitoff projection in equatorial coordinates of the distribution of our program objects on the sky. Red squares show objects with CFHT 
or Clay imaging. Blue squares indicate objects in our expanded (secondary) sample, having structural parameters derived from archival or literature data.
Galactic globular clusters with $R_{\rm GC} < 25$~kpc are shown as gray squares. Galaxies that were not included in our study 
are shown as green squares.
{\it lower panel:} Same as the upper panel except in Galactic coordinates.}
\label{fig:allsky_equatorial}
\end{figure}

Based on the above criteria, we identified in 2009 a total of 41 satellites belonging to the outer halo
of the Milky Way. This sample includes a mixture of globular clusters \citep{harris96a,harris10a}
and classical or ultra-faint dwarf galaxies (see, e.g., the compendia of \citealt{mateo98a} and 
\citealt{mcconnachie12}). Three massive satellites --- the LMC, SMC and Sagittarius dwarf spheroidal galaxy --- 
were deemed too large and luminous to observe efficiently within the context of this survey and were 
excluded. At the same time, two satellites that were discovered after our survey began --- {\tt Segue~3} and 
{\tt Pisces~II} \citep{belokurov10a} --- were added to our sample during the 2010A semester. A third object residing in
the outer halo --- the ultra-faint globular cluster {\tt Mu{\~n}oz~1}, which is offset by $\sim 1\fdg8$ from the center
of the {\tt Ursa~Minor} dwarf galaxy but located well in the foreground  --- was, in fact, discovered serendipitously 
in this survey \citep{munoz12a}. Thus, our primary sample consists of 44 objects that, aside from the
aforementioned cases of the LMC, SMC and Sagittarius, represents a complete sample of
outer halo satellites as of 2010.

Table~\ref{tab:samp1} gives some basic information for these 44 objects. From left to right, the columns
record an ID number, adopted name, other names appearing in the literature, position in  equatorial ($\alpha$,~$\delta$) and 
Galactic ($l$,~$b$) coordinates, reddening $E(B-V)$ from \citet{schlafly11a}, Galactocentric distance and Cartesian coordinates in a 
Galactocentric frame,
\begin{equation}
\begin{array}{lcl}
X & = & R_{\odot}\cos{b}\cos{l} - X_{\odot}\\
Y & = & R_{\odot}\cos{b}\sin{l}\\
Z & = & R_{\odot}\sin{b},
\end{array}
\label{eq1}
\end{equation}
where $X_{\odot} = 8.5$~kpc is our adopted distance to the Galactic center, $R_{\odot}$ is the heliocentric
distance to each object, so that $R_{\rm GC} = \sqrt{X^2 + Y^2 + Z^2}$ is the distance to the Galactic center. For each 
satellite, references to the original discovery paper or papers are given in the final column.

As described in \S\ref{sec:secondary}, this sample has been supplemented by an additional 14 satellites discovered 
between mid-2010, when data acquisition for our primary sample was completed, and 2015. This brings our total sample to
58 objects which, in 2015, represented $\simeq$ 95\% of all 
known Milky Way satellites beyond  25~kpc (i.e., excluding the three massive systems).

Figure~\ref{fig:xyz} shows three projections ($YZ$, $XZ$ and $XY$) of our sample in Cartesian coordinates
centered on the Milky Way center (equation~\ref{eq1}).  Objects belonging to our primary and secondary samples
are shown as red and blue crosses, respectively. Globular clusters at $R_{\rm GC}<25$\,kpc are denoted by gray squares, and satellites absent
from our study are shown as green crosses. The inner and outer circles
plotted in each panel show our adopted boundary for the outer halo and the virial radius of the Milky Way
according to \citet{dehnen06}, respectively. The arrow indicates the direction to M31 in each projection.
Figure~\ref{fig:allsky_equatorial} shows the distribution of these same populations on the sky, in equatorial 
and Galactic coordinates (upper and lower panel, respectively). 

\subsection{Secondary Sample}
\label{sec:secondary}

Data acquisition for the 44 objects belonging to our primary sample was completed in mid-2011. Between that time
and 2015, a number of new Galactic satellites were identified, many located beyond the Galactocentric
distance of $R_{\rm GC} = 25$\,kpc  that defines the boundary of the outer halo sample. We therefore define a ``secondary"
sample for our survey that is listed in Table~\ref{tab:samp2}. This table presents the same basic information for the
14 new Milky Way satellites as Table~\ref{tab:samp1} did for our primary sample. From left to right, the columns of this
table give an ID number, adopted name, other names appearing in the literature, position in  equatorial and 
Galactic coordinates, reddening from \citet{schlafly11a}, Galactocentric distance, Cartesian coordinates in a 
Galactocentric frame (equation~\ref{eq1}) and references to the original discovery paper or papers.

The 14 objects in our secondary sample includes a mixture of probable or confirmed dwarf galaxies, 
probable or confirmed globular clusters, and objects whose classification remain ambiguous at the present time (i.e., dynamical masses and metallicity distribution functions are needed to surmise their true natures). In every case, 
the initial discovery was based on survey data acquired with wide-field imaging telescopes. For instance,  
imaging from Pan-STARRS led to the detection of {\tt Triangulum~II} \citep{laevens15a} while {\tt Crater} was
discovered independently by \citet{laevens14} using Pan-STARRS and by \citet{belokurov14}  using 
VST/ATLAS\footnote{A french amateur astronomer, Pascal Le D\^u also discovered Crater using a small
aperture telescope. He published his results in the french magazine L'Astronomie in January of 2014. The article can
be found at http://www.cielocean.fr/uploads/images/FichiersPDF/L-Astronomie-\_Janvier2014.pdf}. 
Three additional satellites were discovered in SDSS imaging: {\tt Balbinot~1}  \citep{balbinot13}, {\tt Kim~1}
\citep{kimjerjen15a} and {\tt Pegasus~3} \citep{kim15b}. However, it has been the deployment of the Dark Energy
Camera (DECam; \citealt{flaugher15}) on the Blanco telescope that has produced the largest harvest of satellites. 
A total of seven faint stellar systems ---   {\tt Eridanus~3}, {\tt Horologium~I},
{\tt Reticulum~II},  {\tt Eridanus~II},  {\tt Pictoris~I}, {\tt Tucana~2} and  {\tt Phoenix~2} --- were  
identified by two independent groups \citep{bechtol15a,koposov15a} using imaging from the {\it Dark Energy Survey}
(DES; \citealt{diehl14}). An 
eighth DES satellite, {\tt Grus~I}, was discovered by \citet{koposov15a}, and a ninth, {\tt Indus~1}, was  actually
discovered by \citet{kim15a} using DECam imaging obtained as part of the {\it Stromlo Milky Way Satellite Survey}
and later identified in the DES \citep{bechtol15a,koposov15a}.  A tenth object, {\tt Horologium~II}, was identified
by \citet{kimjerjen15b} using DES Y1A1 public data while {\tt Hydra~II} was discovered by \citet{martin15} 
in their DECam {\it Survey of the Magellanic Stellar HIstory} (SMASH).

\begin{figure}
  \includegraphics[width=0.48\textwidth]{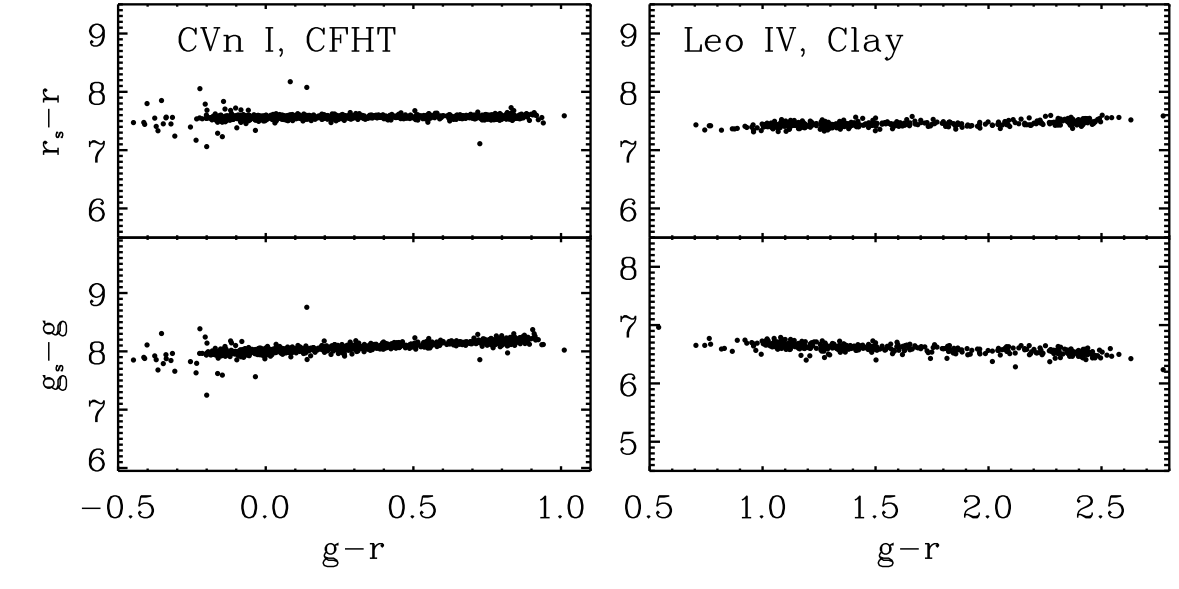}
\caption{({\it Left panels}) Difference between the CFHT instrumental and SDSS calibrated magnitudes as a function of $(g-r)$ 
color for stellar sources in CVn~I. This comparison includes data from all $36$ chips, for objects having  
$g$- and $r$-band SDSS magnitudes in the range 18--21.5. 
({\it Right panels}) Similar to the previous panel, except for stars in Leo~IV as observed by the Clay telescope.}
\label{fig:fig_calib}
\end{figure}

At this stage, the nature of many of these satellites is an area of active investigation. Spectroscopy for member stars
in a handful of systems has provided velocity dispersions, mass-to-light ratios and elemental  abundances for a 
few objects, allowing them to be classified as either dwarf galaxies (e.g., {\tt Horologium~I}, {\tt Reticulum~II}, 
{\tt Hydra~II}; \citealt{koposov15b, simon15,kirby15}) or globular clusters ({\tt Crater}; \citealt{kirby15}). But for most 
of the new satellites, only preliminary classifications are  available --- usually based on their structural properties.

In this survey, we are most concerned with the measurement of structural properties from homogeneous,
high-quality CCD imaging. Fortuitously, for nearly all of these objects, either the discovery or follow-up observations 
includes imaging in the SDSS $g$ and $r$ filters --- the same filter combination used for our primary 
survey. It is therefore possible to maintain the uniformity and homogeneity  of the primary survey by 
adding published photometry, or photometry derived from data in the archive, for these newly discovered 
satellites. Details on the photometric catalogs for our secondary objects --- including both
photometry assembled from the literature and photometry obtained using data retrieved from public 
archives --- will be presented in \S\ref{sec:obs_secondary}.

It is worth noting that the number of known Galactic satellites continues to rise, with many objects identified since 2015 
(e.g., \citealt{kimjerjen15b}, \citealt{kim15b}, \citealt{bechtol15a}, \citealt{koposov15a}, \citealt{luque16a}, \citealt{luque17a}, 
\citealt{laevens15a, laevens15b}, \citealt{torrealba16a, torrealba16b}, \citealt{homma16a}).
At the time of writing,  our sample represents roughly three quarters
of the 77 known satellites having $R_{\rm GC} \ge 25$~kpc. \citet{munoz18a} gives more
details on those satellites that are absent from our samples.

\begin{figure}
  \includegraphics[width=0.485\textwidth]{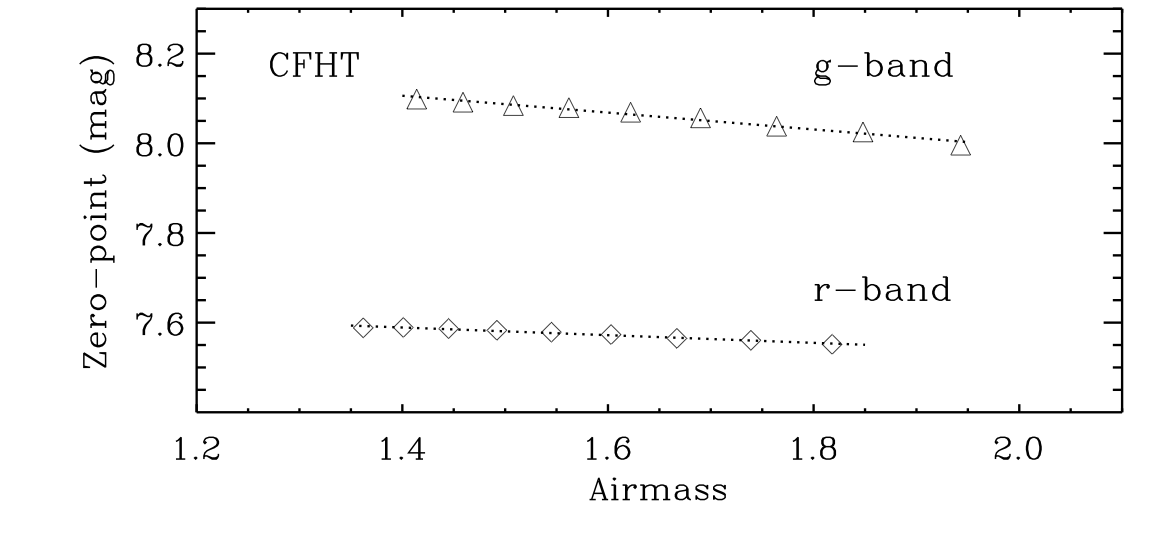}
\caption{
Photometric zeropoint plotted as a function of airmass for CFHT observations of CVn~I.
The upper and lower symbols show the trends observed in the $g$ and $r$ bands, respectively.}
\label{fig:zpoint_am_cfht}
\end{figure}

\section{Imaging and Data Reductions for Primary Targets}
\label{sec:obs_primary}

For our primary sample, observations for northern hemisphere objects were carried out using the MegaCam 
instrument on the 3.6m CFHT. In the south, observations were made using the mosaic camera ---
also named Megacam --- on the 6.5m Magellan II-Clay telescope. To avoid confusion we henceforth refer 
to these instruments as CFHT-MegaCam and Clay-Megacam. Table~\ref{tab:runs} summarizes the details of the 
six CFHT and Clay observing programs, amounting to $\sim$ 100~hrs of telescope time, that comprise our survey. 

\subsection{CFHT-MegaCam Imaging}
\label{subsec:cfht}

CFHT-MegaCam is a wide-field imager consisting of $36$ CCDs --- each measuring $2048\times4612$ pixels --- 
that together cover a $0\fdg96\times0\fdg94$ field of view at a scale of $0\farcs187$~pixel$^{-1}$ \citep{boulade03}. All 
observations were carried out in queue mode during the 2009-A, 2009-B and 2010-A 
observing semesters. Table~\ref{tab:obslog} provides details on the observations for our primary
objects. From left to right, the columns of this table record the target name, telescope, mosaic geometry,
total areal coverage, source of astrometric and photometric calibration (see \S\ref{subsec:astrometry} and
\S\ref{subsec:photometry}, respectively), mean airmass, and
total exposure time in the $g$ and $r$ bandpasses. Although the color baseline offered by these two filters is 
limited, their choice allows us to minimize exposure times for color magnitude diagram (CMD)
analyses (see \S\ref{subsec:comparison2}). Exposure times were chosen so that the $5\sigma$ point-source limiting
magnitude for all objects, and in both bands, lie $\sim$ 2--3 magnitudes below the main sequence turnoff (MSTO).

For 22 of our 30 CFHT targets, a single pointing was adequate to provide complete coverage. For the remaining 
eight objects, a grid of either $2\times1$ or $2\times2$ pointings was used depending
on the spatial extent of the satellite. 
 In all cases, a series of dithered exposures was collected, usually in dark conditions. 
The dithering pattern used was selected from the standard CFHT-MegaCam operation options to provide coverage 
of both the small and large gaps between chips (i.e., the largest vertical gaps in MegaCam 
are six times wider than the small gaps). Typical image quality for the CFHT imaging is $\approx 0\farcs7-0\farcs9$.
Altogether, the CFHT component of our MegaCam survey covers a total area of 43.25 deg$^2$.

For two program objects --- {\tt Pal~3} and {\tt NGC7492} --- imaging was collected using both facilities 
as a cross check on our photometry and astrometry. The results of this comparison will be presented 
in \S\ref{subsec:comparison1} below.

\begin{figure}
  \includegraphics[width=0.48\textwidth]{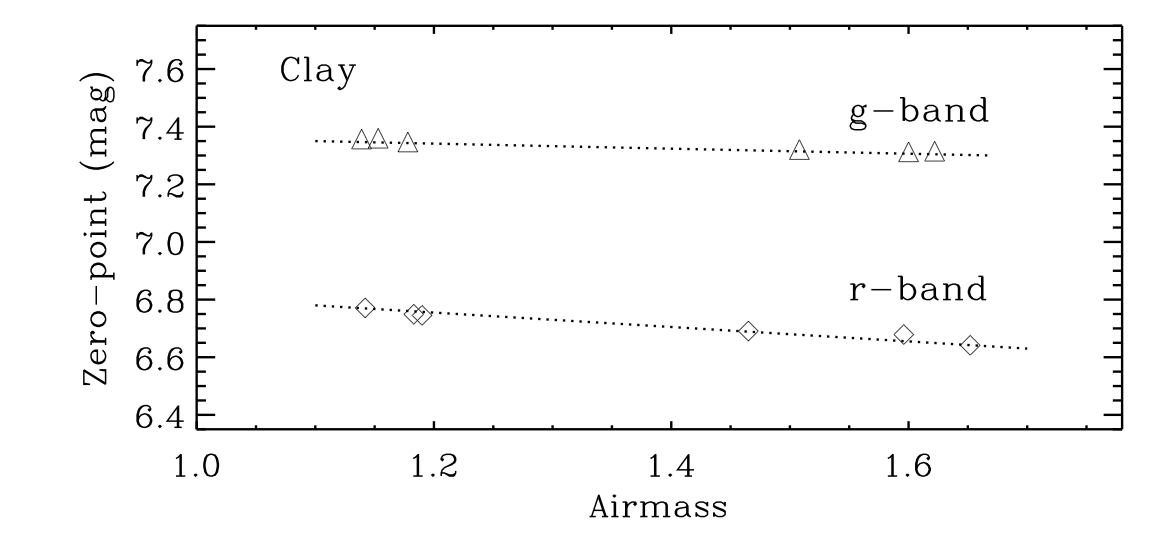}
\caption{
Photometric  zeropoint plotted as a function of airmass for imaging carried out with the Clay telescope. In this
case, the data points show different objects observed using the same exposure time. As in
the previous figure, the upper and lower symbols show the trends in $g$ and $r$ bands, respectively.}
\label{fig:zpoint_am_clay}
\end{figure}

\subsection{Clay-Megacam Imaging}
\label{subsec:clay}

For 16 additional targets, including {\tt Pal~3} and {\tt NGC7492}, in the southern hemisphere, Clay-Megacam imaging 
was acquired during eight nights on the 
6.5m Clay telescope in November 2010 and April 2011. Clay-Megacam is a large mosaic CCD camera that also
consists of $36$ CCDs ($2048\times4608$ pixels) but at a scale of $0\farcs08$~pixel$^{-1}$ \citep{mcleod15}. 
This array provides instantaneous coverage of a $0\fdg4\times0\fdg4$ field. 

Because this field of view is about five times smaller than that of its northern counterpart, we used
multiple pointings for the most extended objects; in the case of Carina, we cover an area of $2.6$\,deg$^2$ in 16 
different fields. For the more compact objects --- usually globular clusters --- only a single pointing
was needed, with the target positioned at the center of the mosaic. To maintain survey homogeneity, 
Clay-Megacam images were also taken in the Sloan $g$ and $r$ filters. In all cases, we collected five 
dithered exposures per pointing in each filter  to  cover chip gaps. Images were usually acquired in dark time and 
during seeing conditions comparable to those at CFHT ($0\farcs7-1\farcs1$). Excluding the two targets
that appear in both our CFHT and Clay programs (see below), the Clay imaging covers a total area of 
8.75 deg$^2$. In all, our MegaCam imaging covers a combined area of 52 deg$^2$.

\subsection{Image Processing and Astrometric Calibration}
\label{subsec:astrometry}

Data from both instruments used in our primary survey were pre-processed prior to
delivery. For  CFHT-MegaCam, preprocessing was done by the CFHT staff using the
standard {\it Elixir} package while the Clay-Megacam data were pre-reduced at the
Harvard-Smithsonian Center for Astrophysics (CfA). In both cases, the goal of preprocessing 
is to provide the user with frames that are corrected for the  instrumental signature across the 
mosaic. This involves bad pixel correction, bias subtraction, flat fielding and the calculation
of preliminary astrometric and photometric solutions that are included in the
headers of the pre-processed images.

However, the World Coordinate System (WCS) information provided with the processed data is only
approximate. For both the Clay and CFHT data, we therefore refine
the astrometric solution using the latest freely available
SCAMP\footnote{{\tt http://astromatic.net/software/scamp/}} package \citep{bertin06a}. 
First, Terapix SExtractor\footnote{{\tt http://astromatic.net/software/sextractor/}} \citep{bertin96a} was run on 
all chips (SCAMP reads in the output files generated by SExtractor) 
and output files were written in the FITS-LDAC format (where LDAC = Leiden Data 
Analysis Center). SCAMP was then run on all chips separately.
SCAMP uses the approximate WCS information in the frames' headers as a starting point, 
and then computes astrometric solutions using external reference catalogs.
In our case, we used the GAIA (DR1, \citealt{gaia16a}) catalog for
39 objects for which the combination of spatial density and magnitude
overlap yielded enough stars in common to determine a reliable solution.
For the other five objects, the solutions based on GAIA were not precise enough due
to the low number of stars per chip in common, and thus we used 
the  SDSS-Data Release 7 (DR7, \citealt{abazajian09a}) for three of them present in SDSS, and
USNO-B1 catalog for the remaining two targets that fall outside the SDSS footprint.
CFHT-MegaCam chips are five times larger in terms of sky coverage than those
of Clay-Megacam so we found systematically more stars in common between
each of our chips and the reference catalog for the CFHT data.
Generally speaking, a couple of hundred stars were used to compute the astrometric solution for
CFHT chips, while several tens of stars were typically used for the case of Clay data.
Despite the difference in sample size, these are sufficient in both cases 
to avoid significant shot-noise
and thus, our astrometric uncertainties do not depend on the
instrument but on the reference catalog used for each object. For objects 
where GAIA was used, we typically obtained global astrometric uncertainties of  
$rms\sim0\farcs04-0\farcs06$ with internal accuracy typically better than$\sim0\farcs02$. 
For those in SDSS, we obtained $rms\sim0\farcs10-0\farcs20$ and for those objects 
in which we used USNO-B1 as the reference catalog,
typical $rms$ uncertainties in the astrometry were $\sim0\farcs3$.

The output from SCAMP is a single FITS header file per processed frame. For the CFHT-MegaCam images, 
this SCAMP output was used to update the WCS information for each chip. Point source photometry 
was then performed on the images with the updated headers (see \S \ref{subsec:photometry}). The final
photometry was then used to translate $x$ and $y$ stellar positions into equatorial coordinates
using the astrometric solution coefficients in the image headers.

In the case of the Clay-Megacam images, the celestial projection used by the CfA team to determine 
the preliminary astrometric solution is zenithal polynomial.  Unfortunately, this projection is incompatible 
with the current version of SCAMP, so the images were reprojected using 
REMAP  into a tangential 
projection (which is SCAMP compatible).  There is, however, an uncertainty involved in translating a given 
pixel value from one projection to another --- a process that introduces small but noticeable differences in 
the magnitudes obtained in the subsequent photometry. Therefore, we 
carried out the photometry in the images with the original zenithal polynomial projection.
The $x$ and $y$ positions of stars in the catalogs were then translated into $x'$ and $y'$ positions
corresponding to the same star in the image reprojected into a tangential projection.
Finally, these tangential $x'$ and $y'$ positions were transformed into equatorial coordinates
using the WCS information obtained using SCAMP in the tangential reprojected image.

\subsection{PSF Photometry}
\label{subsec:photometry}

The photometric processing was similar for images from both telescopes.
Prior to carrying out point-source photometry on our data, we split each mosaic frame
into its 36 individual chips. We then performed point spread function (PSF) photometry by first running DAOPHOT/ALLSTAR
on the individual (non-coadded) frames and then running ALLFRAME on the resulting files,
as detailed in \citet{stetson94a}.

ALLFRAME performs photometry simultaneously on all $g$ and $r$ frames for a given field.
DAOPHOT/ALLSTAR must be run prior to ALLFRAME in order to determine PSF solutions for each chip,
and to generate starlists for individual frames. The
optimum starlists that are needed as inputs to ALLFRAME were generated by cross matching the 
DAOPHOT/ALLSTAR results for the individual frames using the DAOMATCH and DAOMASTER 
packages (\citealt{stetson93a}). 
These packages also provide reasonably good estimates of the spatial offsets
between dithered individual exposures necessary to run ALLFRAME.
Final output files from ALLFRAME were then combined into a single master catalog for each program object.

\subsection{Photometric Calibration: Objects in Common with SDSS}
\label{subsec:calibration1}

For objects that fall inside the SDSS footprint, our instrumental magnitudes have been calibrated
through a direct comparison to the SDSS-DR7.
First, we matched our photometric catalog for each object with the SDSS 
stellar catalog, typically finding several hundred stars per chip in common with the SDSS.
To determine zeropoints and color terms, we used only SDSS stars with $18<r_{\rm SDSS} <21.5$ and
$18<g_{\rm SDSS} <22$.
The faint limit was chosen to eliminate stars from SDSS with large photometric uncertainties and
the bright limit was chosen to avoid saturated stars in our MegaCam data.
We then used the matched catalog to fit equations of the form:

\begin{equation}
\begin{array}{rrcll}
g_{\rm SDSS} & = & g + g_{\rm 0} + g_{\rm 1}(g-r) \\
r_{\rm SDSS} & = & r + r_{\rm 0} + r_{\rm 1}(g-r).
\end{array}
\label{eq2}
\end{equation}

\noindent Here $g$ and $r$ are our instrumental magnitudes,  
$g_{\rm 0}$ and $r_{\rm 0}$ are the zeropoints and 
$g_{\rm 1}$ and $r_{\rm 1}$ are the color terms,
Because we are calibrating directly to SDSS photometry, we do not need to determine 
the airmass terms.

In their CFHT-MegaCam study of {\tt Coma Berenices} and {\tt Ursa Major~II},  \citet{munoz10a} 
derived zeropoints and color terms for each chip individually, in order to examine possible 
chip-to-chip variations for this instrument. In both cases, they found that the chip-to-chip differences, for both
the zeropoints and color terms, were smaller than the uncertainties in the derived parameters.
For this study, we repeated this test using {\tt CVn~I} and {\tt Segue~1} and found similar results.
Unfortunately, for the Clay-Megacam data, we could not carry out the same test given the low number
of stars per chip in common with SDSS (i.e., all the Clay targets that fall within the SDSS footprint are 
ultra-faint dwarf galaxies with low stellar counts). As an alternative, we used stars in the overlapping 
regions between chips to assess whether there were systematic chip-to-chip variations.
For these stars, we found that the average magnitude difference between the chips was always 
smaller than the magnitude uncertainties of our stars. For each satellite, we therefore combined stars from
all 36 chips to derive global zeropoint and color term values via a linear least-squares fit (weighting by the 
respective uncertainties in the ALLFRAME magnitudes and rejecting $3\sigma$ outliers).

We calculated zeropoints and color terms for each mosaic field independently. 
For the CFHT-MegaCam calibration, 
uncertainties in the zeropoints were typically $0.003-0.004$~mag.
While the $g'_{\rm 0}$ and $r'_{\rm 0}$ terms are a function of exposure time and airmass, the 
color terms remain fairly constant for all objects, with variations of less than $2$\%.
For the Clay-Megacam calibration, typical uncertainties in the zero points are $0.002-0.009$~mag. 
The color terms showed object-to-object variations of less than $5$\%.
A comparison between the CFHT and SDSS magnitudes used to calibrate our
photometry is shown in Figure~\ref{fig:fig_calib} for two representative stellar systems:
{\tt CVn~I}  (CFHT) and {\tt  Leo~IV} (Clay).

\subsection{Photometric Calibration: Objects Not in Common with SDSS}
\label{subsec:calibration2}

Some objects in our primary sample fall outside the SDSS footprint and therefore require a different method 
of  calibration. In such cases, the instrumental magnitudes were calibrated by applying the following equation:

\begin{equation}
\begin{array}{rrcll}
g_{\rm SDSS} & = & g + g_{\rm 0} + g_{\rm 1}(g-r) + g_{\rm 2}X \\
r_{\rm SDSS} & = & r + r_{\rm 0} + r_{\rm 1}(g-r)+ r_{\rm 2}X.
\end{array}
\label{eq3}
\end{equation}

\noindent where $g_{\rm 2}$ and $r_{\rm 2}$ are the airmass terms, and $X$ is the airmass.

The full set of coefficients $g_{0}$, $g_{1}$, $g_{2}$, $r_{0}$, $r_{1}$ and $r_{2}$
where determined using 
the photometry of objects in SDSS as secondary standards. 
The color terms derived for CFHT were
$\langle{g_{\rm 1,CFHT}}\rangle=0.203\pm0.003$ and $\langle{r_{\rm 1,CFHT}}\rangle=0.021\pm0.002$, while
for Clay, we obtained $\langle{g_{\rm 1,Clay}}\rangle=-0.098\pm0.010$ and $\langle{r_{\rm 1,Clay}}\rangle=0.052\pm0.005$.

To determine the airmass terms, we calculated the variation of the zeropoints as a 
function of airmass for objects in the SDSS.
Figure~\ref{fig:zpoint_am_cfht} 
shows the $g$- and $r$-band zeropoints obtained for a variety of airmasses and the linear 
trends fitted to them:

\begin{equation}
\begin{array}{rrcll}
g_{\rm 2,CFHT} & = & -0.176\pm0.015 \\
r_{\rm 2,CFHT} & = & -0.080\pm0.007 
\end{array}
\label{eq5}
\end{equation}

To derive the Clay airmass terms, we used different objects at different airmasses, 
taking advantage of the fact that most of the objects observed at Clay had the same exposure time.
In particular, we used those objects having the lowest photometric errors in the Clay sample: {\tt Leo~IV}, 
{\tt Leo~V}, {\tt Palomar~3}, {\tt Segue~1}, and two different fields for {\tt Leo~II}. 
Figure~\ref{fig:zpoint_am_clay} shows zeropoints obtained for these systems at
different airmasses. The best-fit linear trends yield airmass terms of:

\begin{equation}
\begin{array}{rrcll}
g_{\rm 2,Clay} & = & -0.222\pm0.011 \\
r_{\rm 2,Clay} & = & -0.094\pm0.009. 
\end{array}
\label{eq6}
\end{equation}

Lastly, we determined zeropoints $g_{\rm 0}$ and $r_{\rm 0}$.
For our Clay targets, this was carried out simultaneously with the measurement of the extinction terms since
the exposure times were the same. The resulting zeropoints were found to be:

\begin{equation}
\begin{array}{rrcll}
g_{\rm 0,Clay} & = & 7.016\pm0.014 \\
r_{\rm 0,Clay} & = & 7.463\pm0.011. 
\end{array}
\label{eq7}
\end{equation}

Meanwhile, for the CFHT observations, the zeropoint variations with exposure time (after correcting 
for airmass) had to be computed explicitly since, in this case, the images were taken over a fairly
wide range in exposure time. For objects in common with the SDSS we found: 

\begin{equation}
\begin{array}{rrcll}
g_{\rm 0,CFHT} & = & 1.476 + 2.5\log{T_{exp}} \\
r_{\rm 0,CFHT} & = & 1.014 + 2.5\log{T_{exp}}
\end{array}
\label{eq8}
\end{equation}

Using these relations, zeropoints were then calculated for the remaining objects using their respective 
exposure times. 

Table~\ref{t:photometry_all} presents the full catalogs for all  44 primary targets. The table includes 
the 2000.0 equatorial coordinates, calibrated, unreddened $g$ and $r$ magnitudes as well as 
their uncertainties. We also include the DAOPHOT $chi$ and $sharp$ parameters. We removed the majority of 
spurious and non-stellar detections by applying the following cut: $-0.5<sharp<0.5$ and $chi<3$.
Stars from different objects can be distinguished by their ID name.

\begin{figure}
  \includegraphics[width=0.47\textwidth]{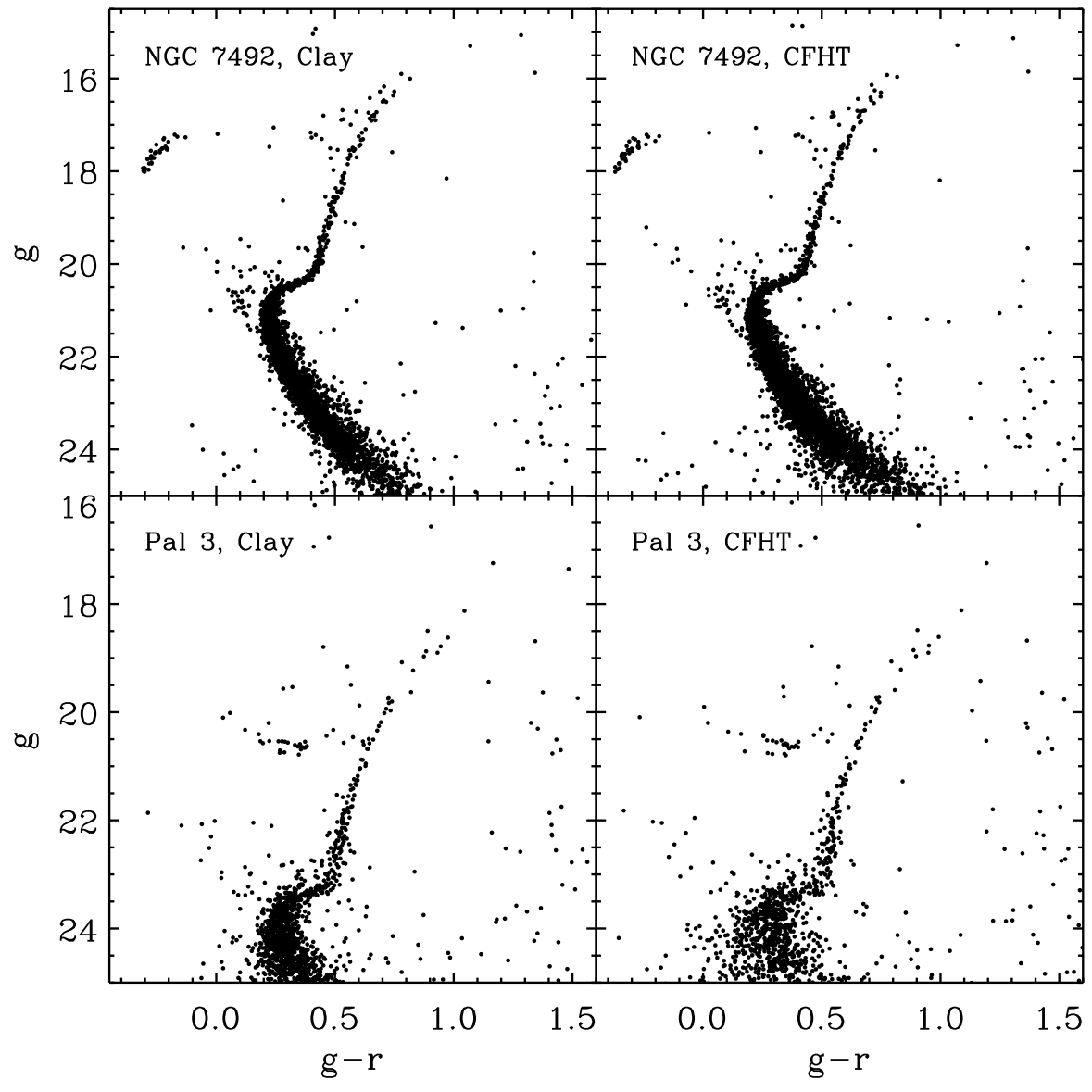}
\caption{
Comparison of the color-magnitude diagrams for two of our program objects, NGC7492 and Pal~3, observed
from both hemispheres. The left and right panels show data within the half-light radii obtained with Clay and CFHT, respectively. The 
Clay-Megacam data appear deeper showing narrower sequences at fainter magnitudes.}
\label{fig:cmd_comp}
\end{figure}

\section{Imaging and Data Reductions for Secondary Targets}
\label{sec:obs_secondary}

As detailed in \S\ref{sec:secondary},
in addition to the 44 objects observed with the Megacam imagers, we include 14 objects discovered after
the completion of our observing campaign. 
For eight satellites discovered using DECam data from the DES survey \citep{bechtol15a,koposov15a}, {\tt Eridanus~II} and {\tt 3}, 
{\tt Horologium~I}, {\tt Reticulum~II}, {\tt Pictoris~I}, {\tt Indus~1},  {\tt Grus~I} and {\tt Phoenix~2}  we retrieved 
archival DECam data from the NOAO Science archive\footnote{http://www.portal-nvo.noao.edu/}. 
In the case of Indus~1, discovered independently, we also obtained Kim's et al. 
photometry. 
We obtained the photometry for two other satellite candidates, {\tt Peg~3} \citep{kim15b} and {\tt Tucana~2} (from the
DES sample), but unfortunately our method to retrieve structural parameters was not able to converge due to the low number 
of stars in the filed and therefore we do not include them in the final list.
Table~\ref{tab:obslog2} presents a summary of observations for the secondary targets.

The archival data used in this catalog consisted, in most cases, of one DECam pointing observed in 
both the $g-$ and $r-$bands. The DECam imager consists of $62$ $2048\times4096$ pixel chips 
with a pixel scale of $0.2626$\,arcsec/pixel covering a total area of $3$\,deg$^{2}$.
In some cases ({\tt Horo~I}, {\tt Ret~II}, {\tt Eri~II}), a second DECam pointing overlapping the 
position of the satellite was present in the archives and thus both pointings were reduced together
and combined to cover the chip gaps. In all cases, the exposure times were $90$ seconds. 
The subsequent photometry procedure was similar to that carried out for the Megacam imagers, i.e., 
DAOPHOT, ALLSTAR was performed in all the individual images  and ALLFRAME was performed in the
cases where more than one observation per field was used. 
 
 Equatorial coordinates for all objects detected by the DAOPHOT/ALLSTAR routines were obtained using
 the WCS information provided in the image headers. Comparison between stellar detections 
 present in multiple observations
 of the same field showed that the internal precision was better than $0.1$\,arcsec. 
 
 \begin{figure}
 \includegraphics[width=0.48\textwidth]{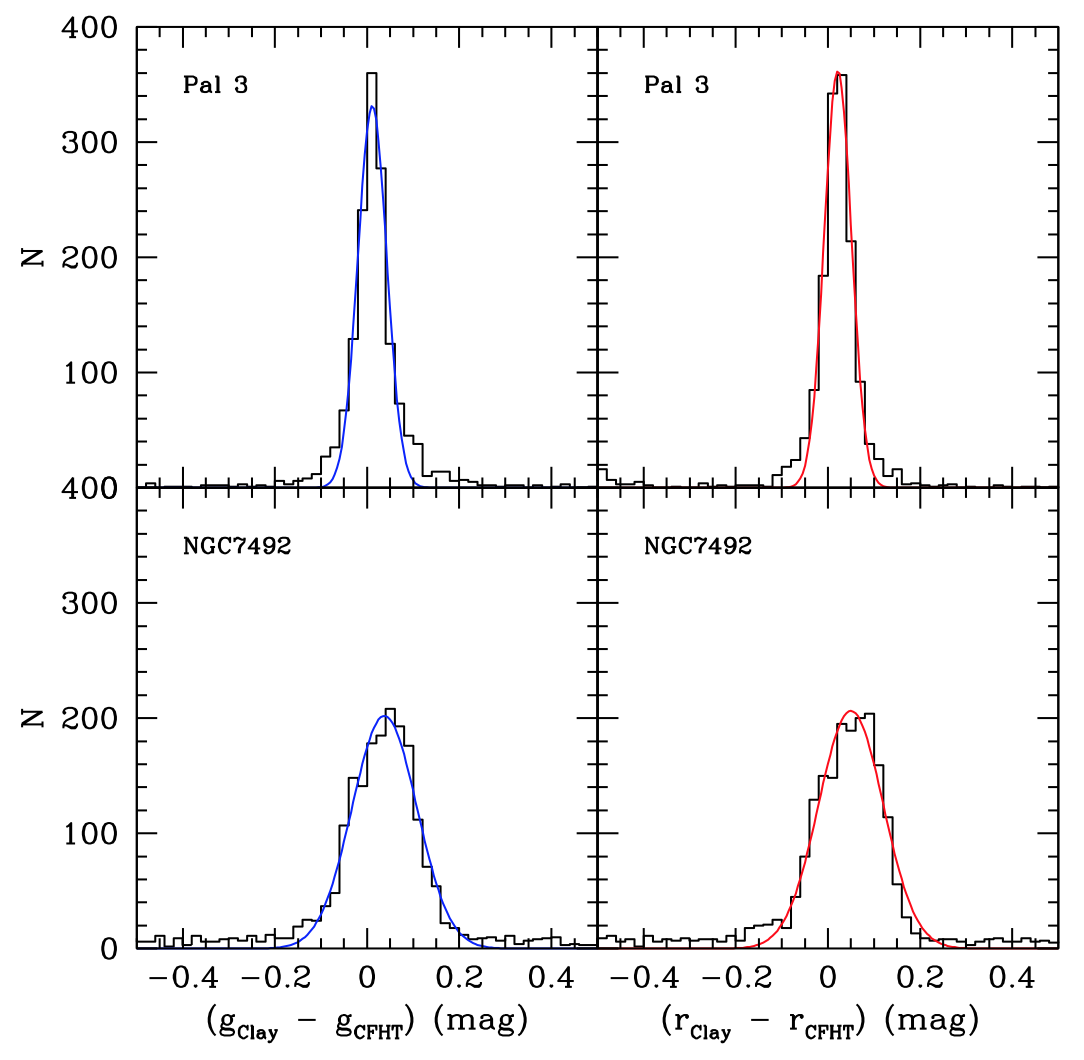}
\caption{
Histograms showing the difference in $g$- and $r$-band magnitude for stellar sources in
Pal~3 (upper panels) and NGC7492 (lower panels), both of which were observed from Clay and CFHT
The smooth curve in each panel shows the best-fit gaussian distribution.} 
\label{fig:hist_comp1}
\end{figure}
 
 To calibrate the instrumental photometry we used DECam data taken for a different program in the same
 bands. From our own DECam data we estimated the zero points and color terms to be:
 
 \begin{equation}
\begin{array}{rrcll}
g_{\rm 0,DECam} & = & 4.960\pm0.031\\
r_{\rm 0,DECam} & = & 5.010\pm0.024\\
 \end{array}
\label{eq9}
\end{equation}
 and
 
 \begin{equation}
\begin{array}{rrcll}
g_{\rm 1,DECam} & = & 0.102\pm0.026\\ 
r_{\rm 1,DECam} & = & 0.113\pm0.021\\
  \end{array}
\label{eq10}
\end{equation}

These values are consistent within the uncertainties with the zero points derived by the 
DECam SMASH survey of the Magellanic Clouds \citep{nidever17a}.

\begin{figure*}
 \plotone{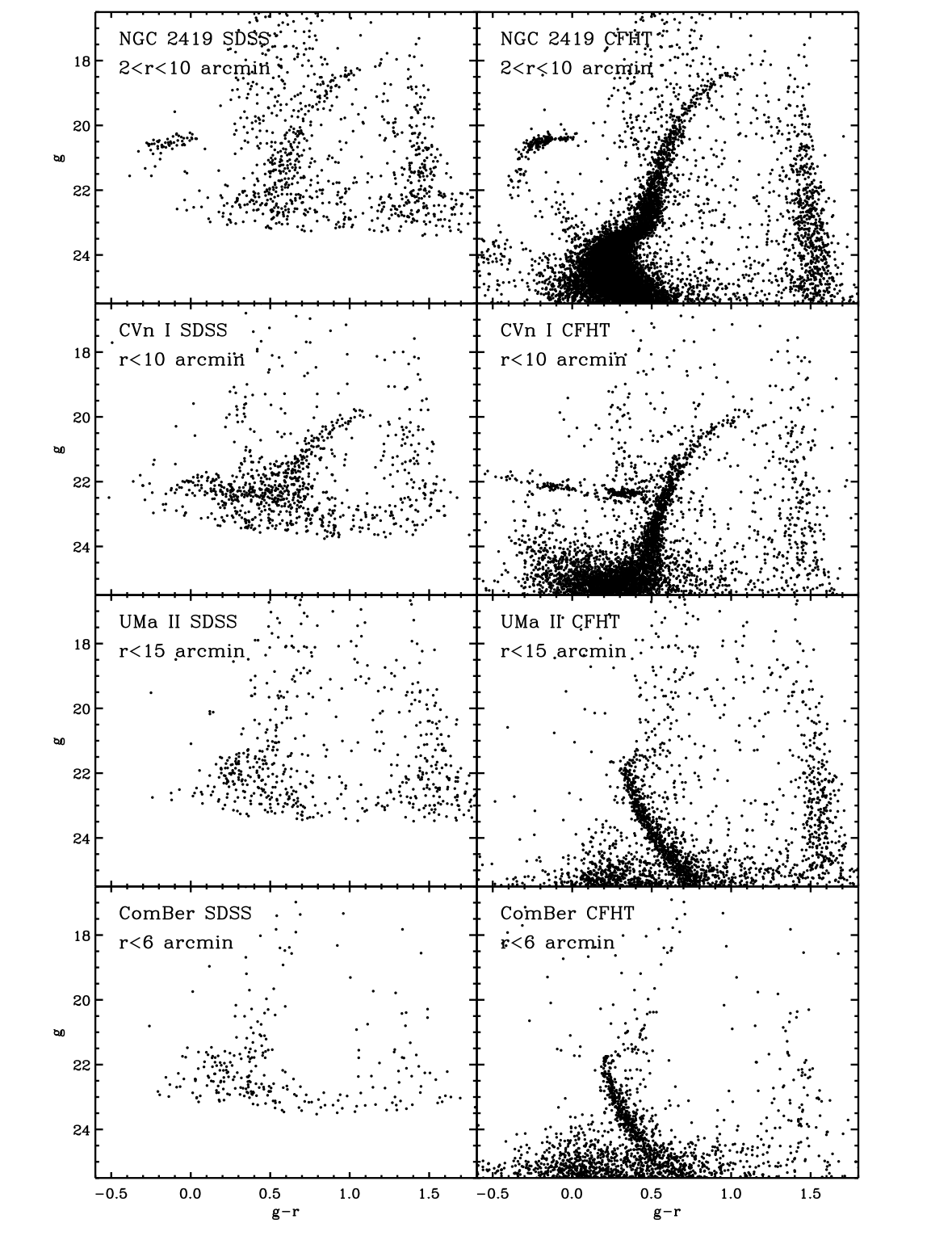}
\caption{
Color-magnitude diagrams for four of our program objects: NGC2419, CVn~I, UMa~II and ComBer. The panels on the
left show CMDs based on data from SDSS (DR12) while those on the right show our new CFHT and Clay photometry.}
\label{fig:comp_sdss}
\end{figure*}

Unfortunately, we were not able to derive an airmass term from our dataset, and therefore we used the 
same zero-points for all the DECam data we processed. If we assume that the missing airmass term 
is similar to those derived for Clay and CFHT the uncertainty introduced in the zero-points by not correcting 
for this effect is of the order of $0.05$ magnitudes in the $g-$band and $0.02$ magnitudes in the
$r-$band. We include these values when estimating the global photometric uncertainties and the 
subsequent luminosity values derived from them.

For an additional six satellites, {\tt Laevens~1} \citep{laevens14}, also
known as {\tt Crater} \citep{belokurov14}, {\tt Triangulum~II} 
\citep{laevens15a}, {\tt Horologium~II} \citep{kimjerjen15b}, {\tt Hydra~II} \citep{martin15},  {\tt Kim~1} \citep{kimjerjen15a} and
{\tt Kim~2} \citep{kim15a} 
and 
{\tt Balbinot~1} \citep{balbinot13}, the respective authors were kind enough to send us their
photometric catalogs for the purpose of measuring their structural parameters. 

\begin{figure}
  \includegraphics[width=0.48\textwidth]{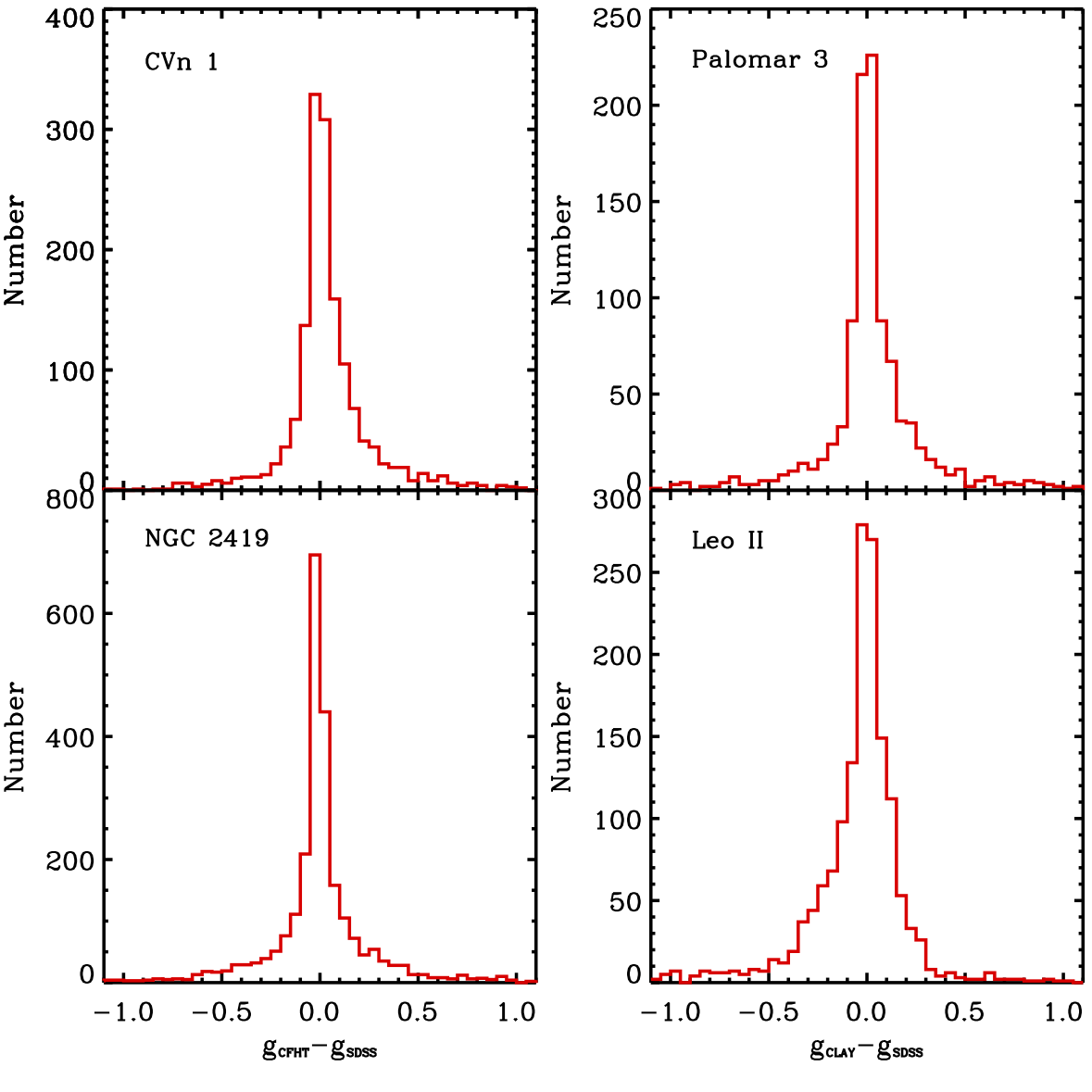}
\caption{
({\it Upper left panel}) Histogram of the difference between the $g-$ band magnitude from the CFHT and the SDSS catalogs for {\tt CVn~1}.
({\it Lower left panel}) Same as above but for {\tt NGC2419}.
({\it Upper right panel}) Histogram of the difference between the $g-$ band magnitude from the Clay and the SDSS catalogs for {\tt Palomar~3}.
({\it Lower right panel}) Same as above but for {\tt Leo~II}.}
\label{fig:phot_comp_g}
\end{figure}

\section{Consistency Checks}
\label{sec:checks}

\subsection{Comparison of CFHT and Clay Photometry}
\label{subsec:comparison1}

Three objects in our primary sample were observed using {\it both} CFHT and Clay:
{\tt Palomar~3}, {\tt Segue~1} and {\tt NGC7492}. This provides us with an opportunity to assess the overall 
homogeneity of the photometry obtained with the northern and southern facilities.
Although the exposure times were slightly shorter for our Clay observations (see Table~\ref{tab:obslog} for 
details) this is roughly offset by the larger telescope aperture, resulting in comparable depths in both
objects. Because a single pointing was used for both targets, though, the CFHT data have the advantage
of covering a $\sim$ five times larger field.

The upper and lower panels of Figure~\ref{fig:cmd_comp} show CMDs for the inner 
regions of {\tt Palomar~3} and {\tt NGC7492}, respectively. Results from Clay are shown in the left panels, while
those from CFHT are shown on the right. 
A visual inspection of this figure shows that the data are of comparable quality
at the bright end, but the Clay-Megacam data are deeper, 
which is evident by the narrower main sequence at the faint end.
To compare the photometry  
from the two instruments, Figure~\ref{fig:hist_comp1} shows histograms for
$(g_{\rm Clay}-g_{\rm CFHT})$ and $(r_{\rm Clay}-r_{\rm CFHT})$ for both objects. This comparison
uses sources down to $r_{\rm Clay}=24$ and applies a cut of $-0.5 < {\tt sharp} < 0.5$ to
isolate only star-like detections.
For {\tt Palomar~3}, the distributions are centered on $(g_{\rm Clay}-g_{\rm CFHT})=0.011$ and 
$(r_{\rm Clay}-r_{\rm CFHT})=0.022$ with dispersions of $\sim0.031$ and $\sim0.030$, respectively.
In the case of {\tt NGC7492}, the distributions are centered on $(g_{\rm Clay}-g_{\rm CFHT})=0.038$ and 
$(r_{\rm Clay}-r_{\rm CFHT})=0.049$. The respective dispersions are $0.070$ and $0.070$.

\begin{figure}
  \includegraphics[width=0.48\textwidth]{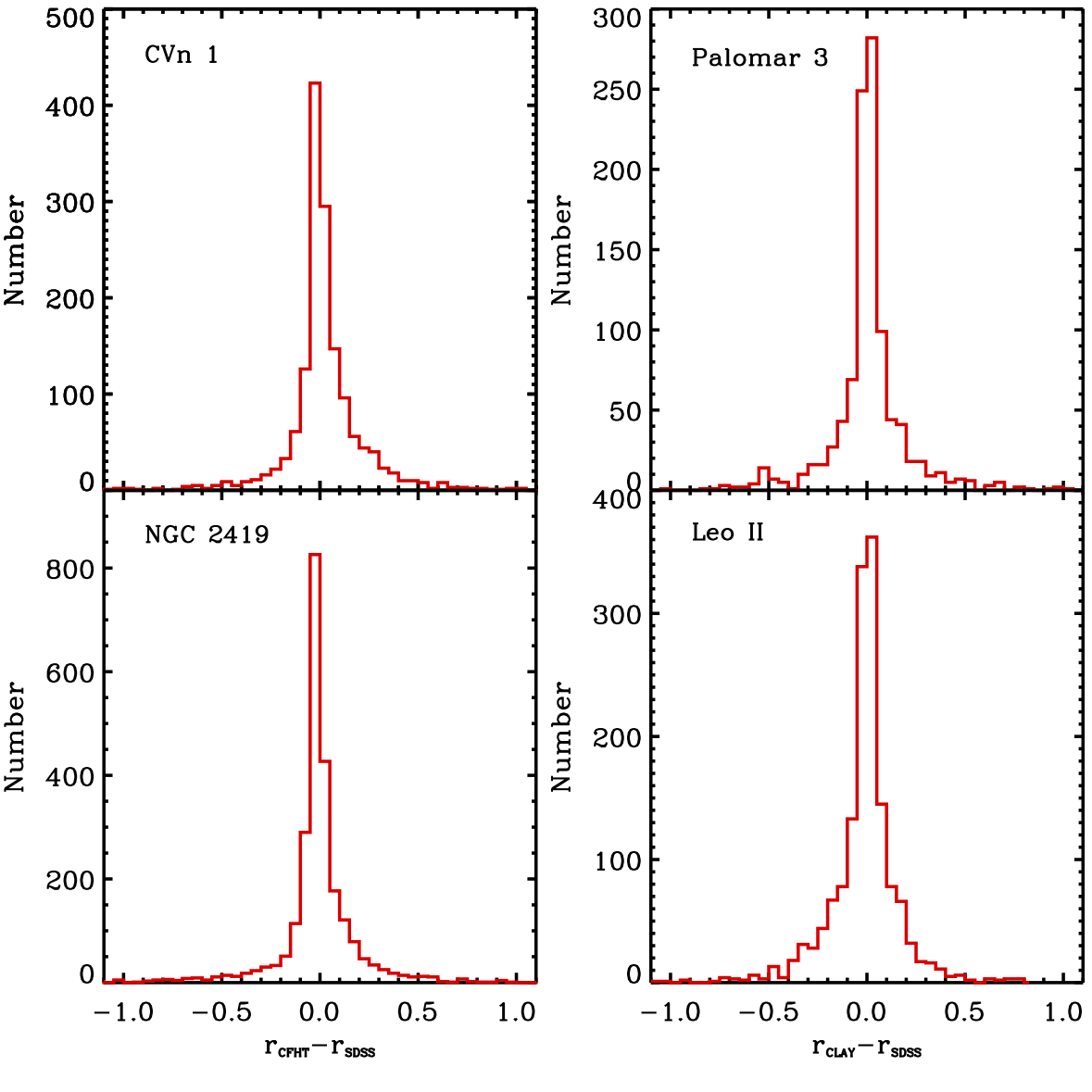}
\caption{
Same as Figure~\ref{fig:phot_comp_g} but for $r-$band magnitudes.}
\label{fig:phot_comp_r}
\end{figure}

\subsection{Comparison of MegaCam and SDSS Photometry}
\label{subsec:comparison2}

Deep, homogenous  photometry for our program objects is essential if we are to achieve the 
scientific goals laid out in \S\ref{sec:intro}: namely, the analysis of CMDs,
star formation histories, density distributions and structural parameters for a nearly complete sample of
outer halo satellites. An obvious point of comparison is the SDSS, which has had an enormous impact
on our census and understanding of the halo and its substructures. Recall from \S\ref{subsec:astrometry} that 
SDSS photometry is available for 27 of our 44 primary survey targets. 

In Figure~\ref{fig:comp_sdss}, we compare our new CMDs to those from
SDSS (DR12) for four of our program objects. From top to bottom, the panels in this figure show
CMDs for {\tt NGC2419}, {\tt CVn~I},  {\tt UMa~II} and {\tt ComBer}. Results from SDSS are shown 
in the left panels, while those from our program (all based on CFHT data) are shown on the right. The 
comparison has been restricted to the inner regions of the targets (roughly within their respective 
effective radii, the actual radii are shown in the Figure). In all cases, the SDSS data were 
limited to unresolved sources with photometric uncertainties lower than 
$0.25$~mag in both the $g$ and $r$ bands. Similarly, the CFHT data were
restricted to detections having $-0.5 < {\tt sharp} < 0.5$ and $ {\tt chi} < 3$ in order to eliminate as many 
extended sources as  possible. Similar restrictions on the $g$ and $r$ photometric errors were also applied.

For all four objects, Figure~\ref{fig:comp_sdss} shows there is a dramatic improvement in depth and 
precision compared to  SDSS. The SDSS CMDs typically reach only to a depth of $g \sim 22-23$ which
is adequate only to identify the red giant branch at the distances of {\tt NGC2419} and {\tt CVn~I}. For {\tt UMa~I} 
and {\tt ComBer}, it is just possible to identify their main sequence turnoffs (MSTOs). By contrast, all the major 
evolutionary sequences are easily identified and well defined in the panels on the right, as the CFHT-MegaCam 
photometry reaches several magnitudes below the MSTO.

\begin{figure}
  \includegraphics[width=0.48\textwidth]{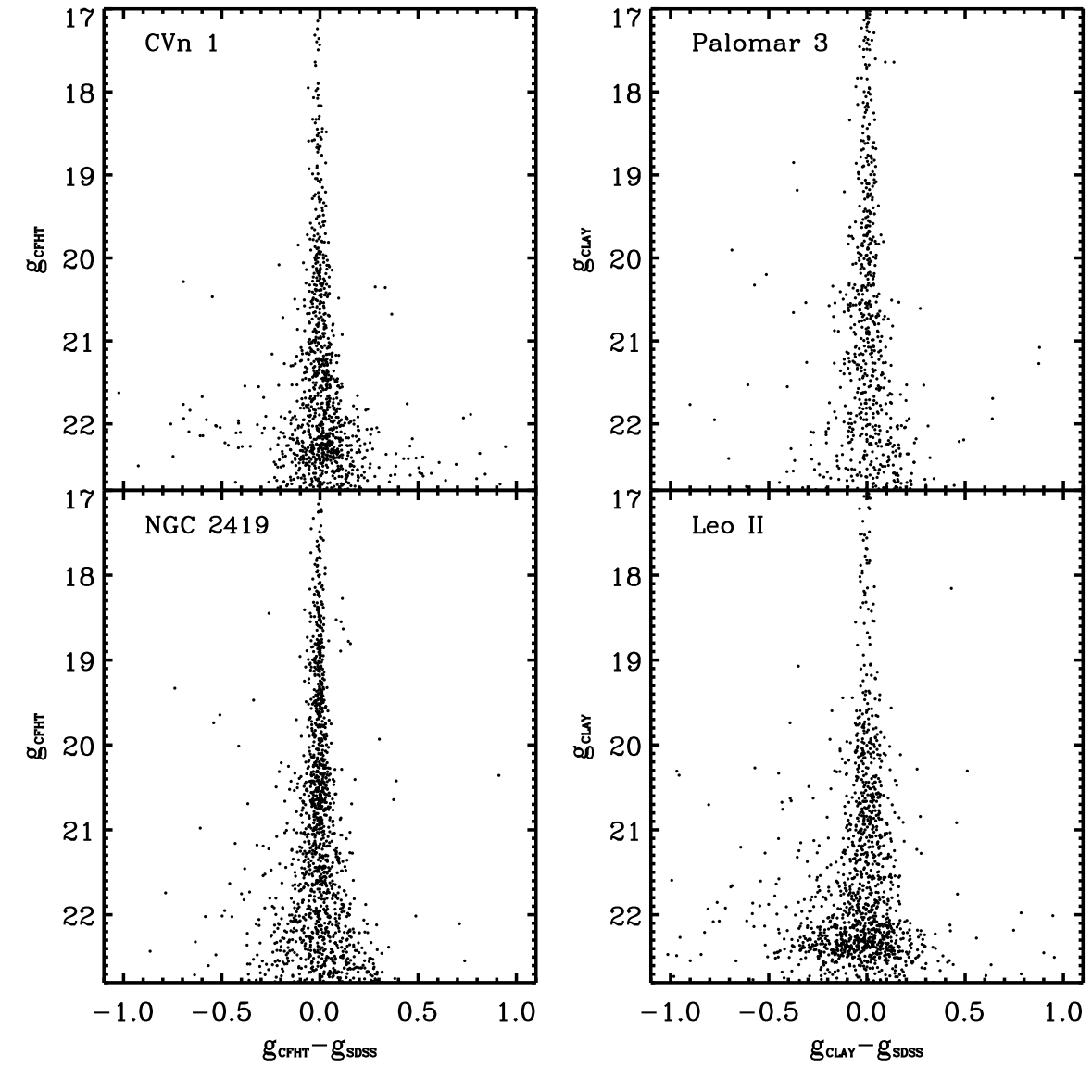}
\caption{
({\it Upper left panel}) Difference in $g-$band magnitudes between CFHT and SDSS data versus CFHT $g-$band magnitude for {\tt CVn~1}. 
({\it Lower left panel}) Same as above but for {\tt NGC2419}.
({\it Upper right panel}) Difference in $g-$band magnitudes between CFHT and SDSS data versus CFHT $g-$band magnitude for {\tt Palomar~3}.
({\it Lower right panel}) Same as above but for {\tt Leo~II}.}
\label{fig:phot_comp_diff}
\end{figure}

Figures~\ref{fig:phot_comp_g} and \ref{fig:phot_comp_r}
show histograms of the difference between our $g-$ and $r-$band magnitudes for a region of $20$\,arcmin
around four objects. The left panels show the difference between CFHT and SDSS data for  {\tt CVn~1} and {\tt NGC2419}
while the right panels show the difference between Clay and SDSS data for {\tt Palomar~3} and {\tt Leo~II}.
Figure~\ref{fig:phot_comp_diff} show the difference in $g-$ magnitudes as a function of depth.
All histograms are centered around zero, as expected, and their dispersions are consistent with the
photometric uncertainties, typically larger for the shallower SDSS data.

\begin{figure}
  \includegraphics[width=0.48\textwidth]{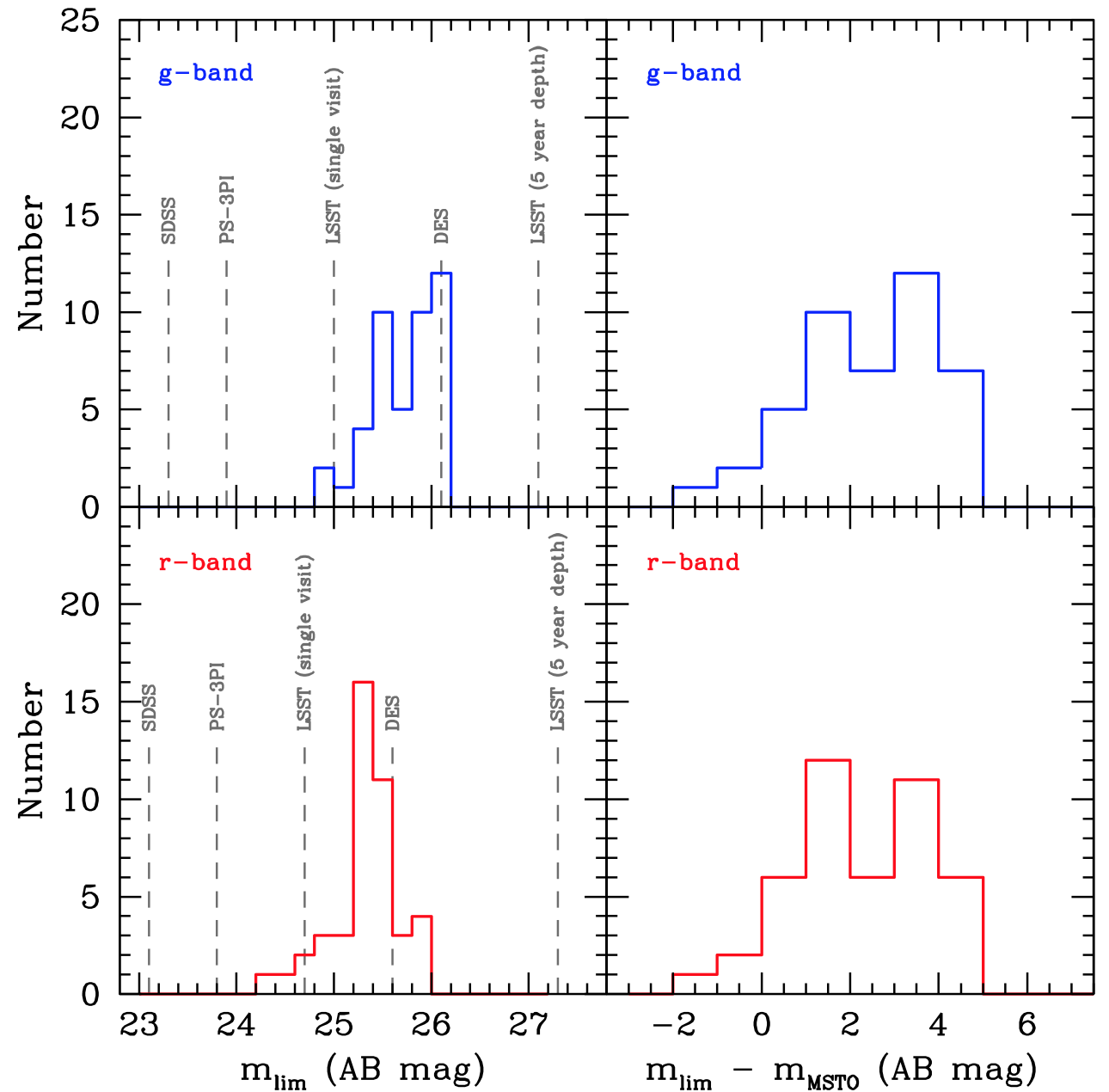}
\caption{({\it Upper left panel}) Distribution of limiting magnitude ($5\sigma$, point source limits) for our 44 CFHT or 
Clay program  objects. For comparison, the dashed vertical lines show the corresponding limits from a number of other 
notable surveys: i.e., Sloan Digital Sky Survey (SDSS), the Pan-STARRS 3PI survey, a single visit from LSST, 
the coadded Dark Energy Survey (DES) and the five year, coadded depth from LSST.
({\it Lower left panel}) Same as above, except for the $r$ band.
({\it Upper right panel}) Difference between our limiting magnitude the main-sequence turnoff magnitude, $m_{\rm MSTO}$,
for the same sample of 44 program objects. Our CFHT and Clay images reach a median depth 
$\simeq$~2.2~mag below the main sequence turnoff in both bands.
({\it Lower right panel}) Same as above, except for the $r$ band.}
\label{fig:depth}
\end{figure}

As noted in \S\ref{sec:obs_primary}, exposure times were chosen to ensure that our photometry would reach 
several magnitudes below the MSTO point in each of our program objects, irrespective of distance. This was
achieved for most objects, except {\tt Leo~T}, {\tt CVn~I} and {\tt NGC2419}.  Figure~\ref{fig:depth} 
illustrates the depths reached in our primary survey. Results for the $g$ and $r-$bands are shown in the 
upper and lower panels, respectively. The panels on the left show the approximate $5\sigma$ limits for
point sources in our MegaCam targets; there is a distribution in limiting magnitude that broadly peaks between
$\sim$ 25 and 26 AB mag. In the right panels, we combine these limiting magnitudes with distances for our
targets to show the approximate depths we reach below the MSTO. As expected, we see broad distributions 
in both bands that peak $\sim$ 2--3 magnitudes below the MSTO.

\begin{figure}
  \includegraphics[width=0.48\textwidth]{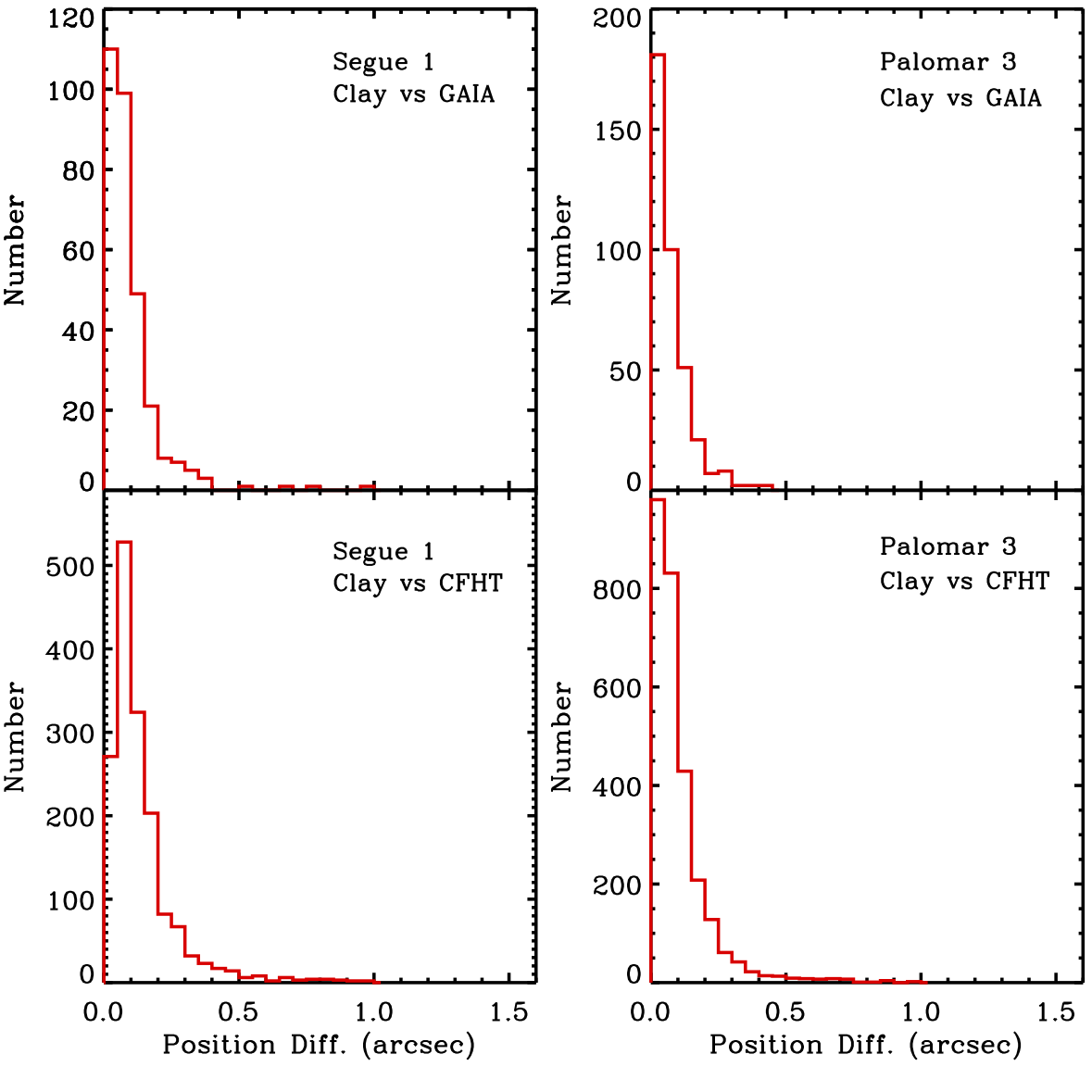}
\caption{
({\it Upper left panel}) Histogram of the difference between the equatorial positions of the same objects in the {\tt Segue~1} field taken 
from the Clay and GAIA catalogs. 
({\it Lower left panel}) Same as above but this time for objects from the Clay and CFHT datasets.
({\it Upper right panel}) Same as {\it upper left panel} but for {\tt Palomar~3}.
({\it Lower right panel}) Same as above but for {\tt Palomar~3}.}
\label{fig:comp_astro}
\end{figure}

Finally, in Figure~\ref{fig:comp_astro} we compare the equatorial coordinates of the stellar-like detections 
($-0.5<sharp<0.5$)
for two objects observed by both imagers,
{\tt Segue~1} and {\tt Palomar~3}. The upper panels show histograms of the difference, in arcsec, 
between the positions determined from 
the Clay and the GAIA data for the same stars. 
The lower panels show the same differences but this time between 
Clay and CFHT data to check for internal consistency. We note that stellar-like objects
down to our magnitude limit are included.

\section{Results}
\label{sec:results}

As an illustration of serendipitous findings from our catalog, we highlight two results.

\subsection{Sagittarius debris in the fields of {\tt Whiting~1} and {\tt NGC7492}}
\label{sec:ngc7492}

One of the advantages of having relatively extended spatial coverage of our Galactic satellites is the 
possibility of studying their outer structure. This is particularly true in the case
of globular clusters observed with CFHT, for which we typically cover several times their half-light (or 
equivalently their effective) radii. For a number of these clusters, secondary main sequences are observed 
beyond the cluster's extent. In some cases, such as {\tt NGC2419} and {\tt Kop~2},  the extra main sequences 
lie in the background or foreground of the clusters \citep{carballobello15}. In other cases, like the ones presented 
here, the sequences seem to be at the same distance of the clusters, indicating either the presence of an 
extended structure related to the cluster (i.e., tidal tails or extended halos) or revealing the existence of a stream 
within which the clusters lie. 

\begin{figure}
  \includegraphics[width=0.49\textwidth]{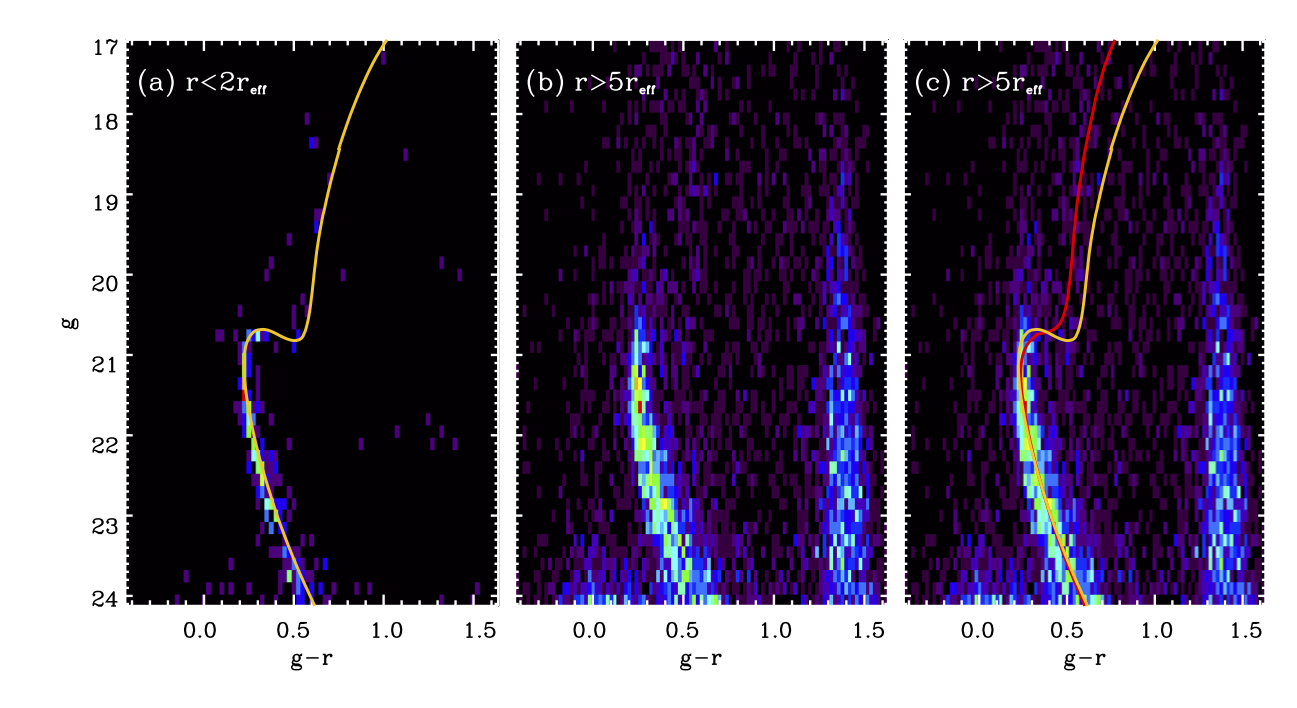}
\caption{
(a) Hess diagram for the inner regions of {\tt Whiting~1}. (b) Hess diagram for the region outside five effective radii.
(c) Same as (b) but with the best-fit isochrone overlaid, indicating that the secondary main sequence
is at a similar distance as the cluster.}
\label{fig:whiting1}
\end{figure}

{\tt Whiting~1} lies at $\sim30$\,kpc from the Sun and is one of the youngest Galactic globular clusters known 
to date with an estimated age of $\sim6.5$\,Gyr, according to \citet{carraro07a}. These authors also 
showed that {\tt Whiting~1}'s distance, position in the sky and mean heliocentric radial velocity ($v_h=131$\,km s$^{-1}$)
coincide almost exactly with the position and radial velocity
of tidal debris from the {\tt Sgr} dSph \citep{majewski03a,law05a}.  They thus
argue that the cluster's origin is most likely to be associated with this dwarf galaxy. 
Through the analysis of $N-$body simulations of the {\tt Sgr} dSph+tail system, \citet{law10a} also associate 
{\tt Whiting~1} with {\tt Sgr}. 
Figure~\ref{fig:whiting1} shows the Hess diagram for the region within two effective radii of the cluster 
(left panel) using our photometry. The cluster's main and scarcely populated evolved sequences are evident. 
Also shown in this panel is the best-fit Dartmouth isochrone \citep{dotter08a}: a $6.5$\,Gyr, [Fe/H]$=-0.7$ track
located at  $d_{\rm helio}=30.5$\,kpc. The middle panel of this figure shows the same diagram but for a region 
beyond five effective radii, where the cluster stellar density is extremely low and thus the contribution 
of cluster stars should be negligible. A secondary main sequence is clearly visible in this region.  
Based on the results from \citet{carraro07a} and \citet{law10a} we associate this ``extra" population
with debris from the {\tt Sgr} galaxy. 

According to the Law et al.'s model, the {\tt Sgr} population at the position
of {\tt Whiting~1} should be a combination of both old trailing and younger leading-arm debris, where the
age refers to the time when the stars became unbound to the galaxy. 
For reference, in the right panel of Figure~\ref{fig:whiting1} we overplot both a [Fe/H]$=-1.4$, $12$\,Gyr old 
and a [Fe/H]$=-0.5$, $6.5$\, Gyr old isochrone located at 
distances of $d_{hel}=26$ and $30.5$\,kpc respectively. These isochrones are meant to represent old and 
intermediate-age {\tt Sgr} populations. The choice of age and metallicity is motivated by the metallicity and age 
distribution of {\tt Sgr} stars along its tails reported by \citet{chou07a} and the star formation history derived by 
\citet{siegel07a}.
Given the lack of detectable stars in the sub- and red-giant branch region of the CMD, with our current
data we cannot discriminate between a young population located at a similar distance as
the cluster and an old but slightly closer one. However, we consider it likely that this
{\tt Sgr} population is a combination of ages and metallicities, and that it spreads in distance over
a range of a few kpc.
 
\begin{figure}
  \includegraphics[width=0.49\textwidth]{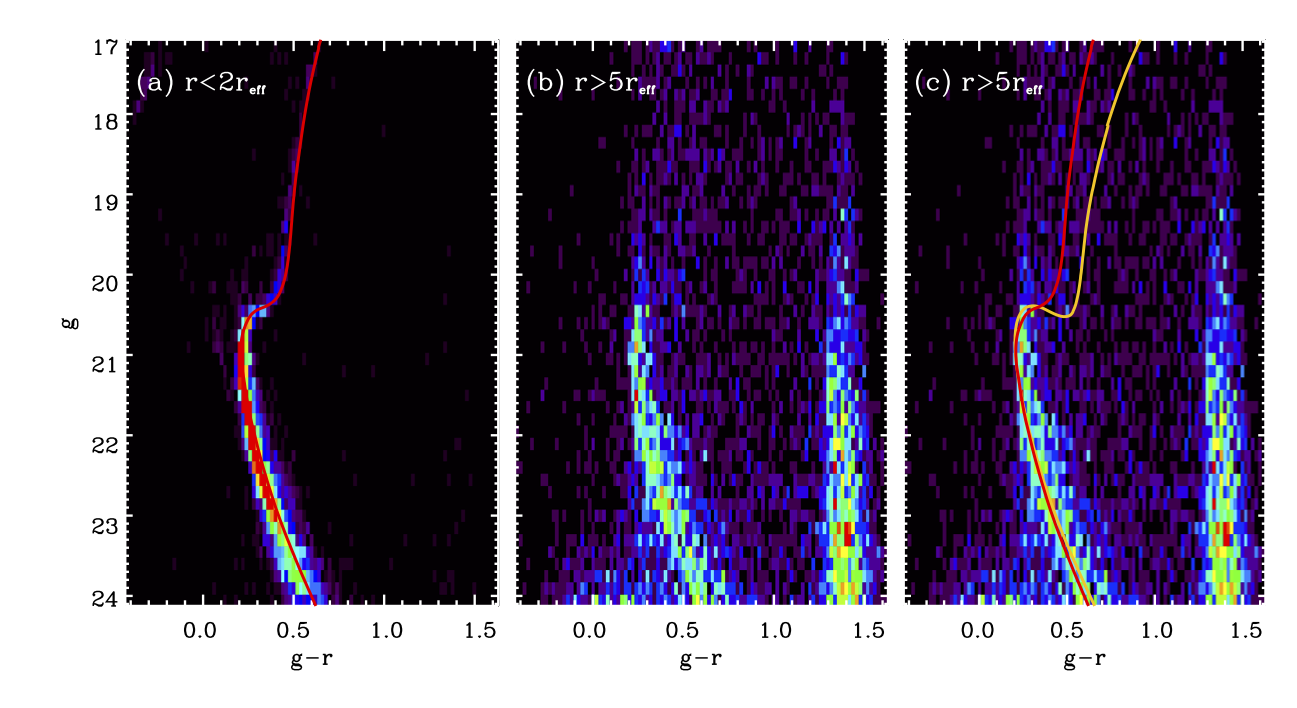}
\caption{
(a) Hess diagram for the inner regions of {\tt NGC7492}. (b) Hess diagram for the region outside five effective radii.
(c) Same as (b) but with the best-fit isochrone overlaid, indicating that the secondary main sequence
is at a similar distance as the cluster.}
\label{fig:ngc7492}
\end{figure}

Another globular cluster located close to tidal debris from the {\tt Sgr} dSph is {\tt NGC7492}.
In this case, \citet{law10a} found a coincidence between the spatial position
of the cluster and its distance with those of {\tt Sgr} debris but 
considered a connection between the cluster and the galaxy unlikely based on the difference in mean 
radial velocity between the two systems.
Figure~\ref{fig:ngc7492} is the equivalent of \ref{fig:whiting1} but this time showing Hess diagrams for our
data in the field of the {\tt NGC7492}.
The left panel again shows the central region of the cluster with the best-fit Dartmouth isochrones, a 
$13$\,Gyr, [Fe/H]$=-1.7$ sequence located at $d_{\rm helio}=24$\,kpc.
The middle panel shows a diagram for the outer regions of the cluster. Here, again, a secondary main 
sequence is clearly visible.  Finally, the right panel is the same as the middle one but 
with the same set of isochrones shown in the right panel of Figure \ref{fig:whiting1} (c) overplotted. The
old isochrone is located at the same distance as the cluster, while the younger isochrone is 
placed at $d_{\rm helio}=28$\,kpc. 

\begin{figure}
  \includegraphics[width=0.48\textwidth]{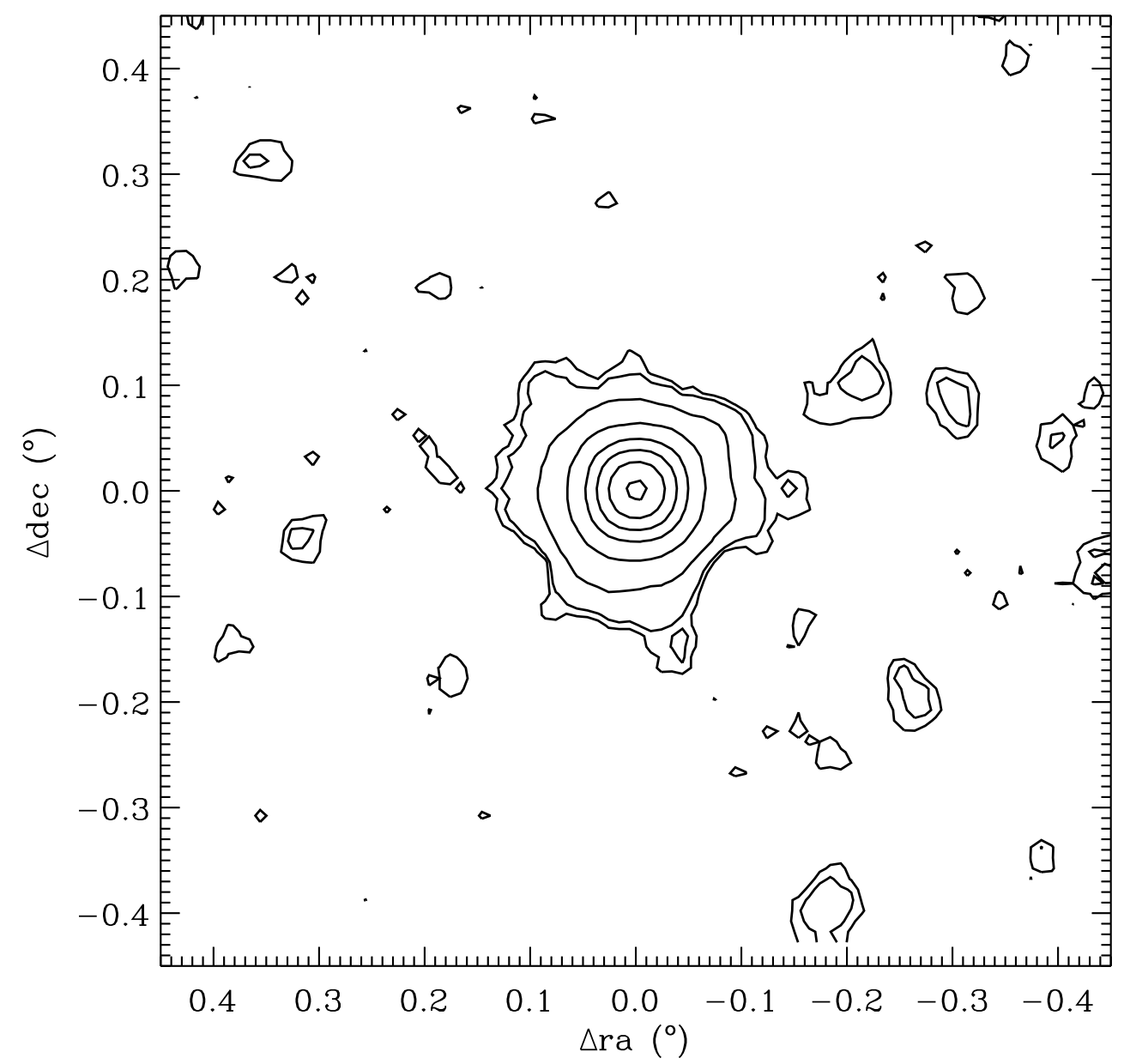}
\caption{
Density contour map for {\tt NGC7492}. The isodensity contours shown correspond to $2, 3, 10, 50, 200, 450, 850$ and $1450$\,$\sigma$ over
the background level. }
\label{fig:ngc7492_contours}
\end{figure}

In addition to the potential connection to {\tt Sgr} debris, it has been reported that {\tt NGC7492} shows
signs that it has been affected by Galactic tides.  Using deep photometry, \citet{lee04a} 
reported an elongated stellar distribution along with the presence of
extended material asymmetrically distributed around the cluster (see their Figure 9). 
Our dataset is slightly deeper and covers a larger area and thus we use it to
investigate the potential extended structure of the cluster.
Unfortunately, our results do not corroborate those of \citet{lee04a}. 
Figure~\ref{fig:ngc7492_contours} shows a density contour map for {\tt NGC7492} where no clear 
elongation or tidal extensions are visible. We find an almost circular stellar distribution past the effective radius of
the cluster, and cannot confirm the existence of an extended, asymmetrical structure surrounding the
system. We suggest that the presence of {\tt Sgr} debris at a similar distance as {\tt NGC7492} affects the 
selection of cluster member stars based on color-magnitude diagram filtering, making it difficult
to reach firm conclusions on the extended structure of this object.

\section{Summary}
\label{sec:summary}

We have described a new, systematic imaging survey of satellites belonging to the outer halo of
the Milky Way (i.e., $R_{\rm GC} \ge 25$~kpc). In a point of departure from most previous studies, our sample 
selection has been made with no constraint on morphology or classification.  
Our primary sample is composed of 44 objects for which we have acquired deep, wide-field 
(0.16--4~deg$^2$) $gr$ imaging with the MegaCam instruments on the 3.6m CFHT and the 6.5m 
Magellan/Clay telescopes. The point-source limiting magnitude for our MegaCam imaging is typically 
2--3 magnitudes below the MSTO in these objects in both bands. Collectively, the survey covers an area
of 52 deg$^2$.
 
This sample has been supplemented with published photometry, or photometry derived from archival 
imaging, for 14 objects discovered between 2010 and 2015. Our final sample of 58 objects represents
roughly three quarters of outer halo satellites known as of 2015 (and roughly three quarters of the
currently known satellites). Our photometric catalog has already been 
used in a series of papers on outer halo satellites, their constituent stars, and possible foreground substructures 
\citep{bradford11,munoz12b, santana13, carballobello15, carballobello17}. In a companion paper, we present 
homogeneous photometric and structural parameters for these satellites \citep{munoz18a} and in future papers we
will examine their scaling relations in the context of other stellar systems.

\acknowledgements

The authors thank Dongwon Kim, Helmut Jerjen and Eduardo Balbinot for kindly sharing their
photometric catalogs for several of our secondary targets.
We also thank an anonymous referee for helping us improve this article.
This work was supported in part by the facilities and staff of the Yale University
Faculty of Arts and Sciences High Performance Computing Center.
R.R.M. acknowledges partial support from project 
BASAL PFB-$06$ as well as FONDECYT project N$^{\circ}1170364$. 
M.G. acknowledges support from the National Science Foundation under
award number AST-0908752 and the Alfred P.~Sloan Foundation.
S.G.D. was supported in part by the NSF grants AST-1313422, AST-1413600, AST-1518308, and by the Ajax Foundation.
This work was supported in part by the facilities and staff of the Yale University Faculty 
of Arts and Sciences High Performance Computing Center.

\clearpage

%%%%%%%%%%%%%%%%%%%%%%%%%%%%%%%%%%%%%%%%%%%%%%%%%%%%%%%%%%%%%%%%%%%%%%
%   REFERENCES
%%%%%%%%%%%%%%%%%%%%%%%%%%%%%%%%%%%%%%%%%%%%%%%%%%%%%%%%%%%%%%%%%%%%%%

%%%%%%%%%%%%%%%%%%%%%%%%%%%%%%%%%%%%%%%%%%%%%%%%%%%%%%%%%%%%%%%%%%%%%%
%   TABLES
%%%%%%%%%%%%%%%%%%%%%%%%%%%%%%%%%%%%%%%%%%%%%%%%%%%%%%%%%%%%%%%%%%%%%%

\clearpage

%\begin{landscape}

\begin{deluxetable*}{lrcrrrrcrrrrrl}
\tablewidth{\textwidth}
\tabletypesize{\footnotesize}
\tablecaption{Basic Parameters for Outer Halo Objects: Primary Sample\label{tab:samp1}}
%\tabletypesize{\scriptsize}
%\tablewidth{0pt}
\tablehead{\colhead{No.} &
\colhead{Name} &
\colhead{Other} &
\colhead{$\alpha$(2000)} &
\colhead{$\delta$(2000)} &
\colhead{$l$} &
\colhead{$b$} &
\colhead{$E(B$-$V)$} &
\colhead{$R_{\odot}$} &
\colhead{$R_{\rm GC}$} &
\colhead{$X$} &
\colhead{$Y$} &
\colhead{$Z$} &
\colhead{Discovery$^\dagger$}  \\
\colhead{} &
\colhead{} &
\colhead{} &
\colhead{(deg)} &
\colhead{(deg)} &
\colhead{(deg)} &
\colhead{(deg)} &
\colhead{(mag)} &
\colhead{(kpc)} &
\colhead{(kpc)} &
\colhead{(kpc)} &
\colhead{(kpc)} &
\colhead{(kpc)} &
\colhead{Reference}}
\startdata
~1 & Sculptor  	 & 		& 15.0183 & --33.7186 &  287.6964 &  --83.1523 & 0.016 &  86.0 &  86.1 &   --5.4 &   --9.8 &  --85.4 & 1 \\ 
~2 & Whiting~1 & 		& 30.7372 &  --3.2519 &  161.6161 &  --60.6353 & 0.021 &  30.1 &  34.9 &  --22.4 &     4.7 &  --26.3 & 2 \\
~3 & Segue~2  	& 		& 34.8226 &  +20.1625 &  149.4461 &  --38.1444 & 0.164 &  35.0 &  41.2 &  --32.2 &    14.0 &  --21.6 & 3 \\
~4 & Fornax  	& 		& 39.9583 &  --34.4997 &  237.2382 &  --65.6740 & 0.020 & 147.0 & 149.1 &  --41.4 &  --50.9 & --133.9 & 4 \\
~5 & AM~1  	& E~1 	& 58.7607 &  --49.6153 &  258.3610 &  --48.4700 & 0.007 & 123.3 & 124.7 &  --25.0 &  --80.1 &  --92.3 & 5 \\
~6 & Eridanus  & 		& 66.1853 &  --21.1876 &  218.1069 &  --41.3325 & 0.018 &  90.1 &  95.4 &  --61.7 &  --41.8 &  --59.5 & 6 \\
~7 & Palomar~2  & 		& 71.5248 &   +31.3817 &  170.5302 &   --9.0719 & 1.015 &  27.2 &  35.5 &  --35.0 &     4.4 &   --4.3 & 7,8 \\
~8 & Carina  	& 		& 100.4066 & --50.9593 &  260.1061 &  --22.2194 & 0.053 & 105.0 & 106.7 &  --25.2 &  --95.8 &  --39.7 & 9 \\
~9 & NGC~2419  & 		& 114.5354 &  +38.8819 &  180.3698 &   +25.2416 & 0.052 &  82.6 &  90.4 &  --83.2 &   --0.5 &    35.2 & 10 \\
10 & Koposov~2 & 		& 119.5715 &  +26.2574 &  195.1098 &   +25.5469 & 0.037 &  34.7 &  42.3 &  --38.7 &   --8.2 &    15.0 & 11 \\
11 & UMa~II 	& 		& 132.8726 &  +63.1335 &  152.4603 &   +37.4411 & 0.082 &  32.0 &  38.5 &  --31.0 &    11.8 &    19.4 & 12 \\
12 & Pyxis 	& 		&136.9869 & --37.2266 &  261.3212 &    +6.9915 & 0.282 &  39.4 &  41.5 &  --14.4 &  --38.7 &     4.8 & 13 \\
13 & Leo~T 	& 		& 143.7292 &  +17.0482 &  214.8598 &   +43.6657 & 0.027 & 417.0 & 422.1 & --256.0 & --172.4 &   287.9 & 14 \\
14 & Palomar~3 & UGC 05439 & 151.3823 &   +0.0718 &  240.1409 &   +41.8642 & 0.037 &  92.5 &  96.0 &  --42.8 &  --59.8 &    61.7 & 7,8 \\
15 & Segue~1 & 		& 151.7504 &  +16.0756 &  220.4777 &   +50.4089 & 0.028 &  23.0 &  28.1 &  --19.6 &   --9.5 &    17.7 & 15 \\
16 & Leo~I  & UGC 5470, DDO 74 & 152.1146 &  +12.3059 &  225.9848 &   +49.1100 & 0.031 & 254.0 & 258.0 & --124.0 & --119.6 &   192.0 & 16 \\
17 & Sextans  & 		& 153.2628 &  --1.6133 &  243.4974 &   +42.2736 & 0.041 &  86.0 &  89.2 &  --36.9 &  --57.0 &    57.9 & 17 \\
18 & UMa~I  	& 		& 158.7706 &  +51.9479 &  159.3625 &   +54.4269 & 0.016 &  97.0 & 101.9 &  --61.3 &    19.9 &    78.9 & 18 \\
19 & Willman~I & SDSS J1049+5103 & 162.3436 &  +51.0501 &  158.5730 &   +56.7834 & 0.012 &  38.0 &  43.0 &  --27.9 &     7.6 &    31.8 & 19 \\
20 & Leo~II  	& Leo B, UGC 6253 & 168.3627 &  +22.1529 &  220.1611 &   +67.2251 & 0.014 & 233.0 & 235.7 &  --77.4 &  --58.2 &   214.8 & 16 \\
      &             & DDO 93 & & & & & & & & & & \\
21 & Palomar~4  & UGCA 237 & 172.3180 &  +28.9733 &  202.3122 &   +71.8009 & 0.020 & 108.7 & 111.5 &  --39.9 &  --12.9 &   103.3 & 7,8 \\
22 & Leo~V  	& 		& 172.7857 &   +2.2194 &  261.8565 &   +58.5343 & 0.024 & 178.0 & 178.8 &  --21.7 &  --92.0 &   151.8 & 20 \\
23 & Leo~IV  	& 		& 173.2405 &   --0.5453 &  265.4577 &   +56.5059 & 0.022 & 154.0 & 154.6 &  --15.2 &  --84.7 &   128.4 & 15 \\
24 & Koposov~1  & 		& 179.8253 &  +12.2615 &  260.9700 &   +70.7551 & 0.022 &  48.3 &  49.5 &  --11.0 &  --15.7 &    45.6 & 11 \\
25 & ComBer 	& 		& 186.7454 &  +23.9069 &  241.8645 &   +83.6122 & 0.015 &  44.0 &  45.2 &  --10.8 &   --4.3 &    43.7 & 15 \\
26 & CVn~II  & SDSS J1257+3419 & 194.2927 &  +34.3226 &  113.5747 &   +82.7013 & 0.009 & 160.0 & 160.7 &  --16.6 &    18.6 &   158.7 & 15,21 \\
27 & CVn~I  	& 		& 202.0091 &  +33.5521 &   74.3039 &   +79.8289 & 0.012 & 218.0 & 217.8 &     1.9 &    37.1 &   214.6 & 22 \\
28 & AM~4  	& 		& 209.0883 & --27.1635 &  320.2827 &   +33.5116 & 0.046 &  32.2 &  27.5 &    12.1 &  --17.2 &    17.8 & 23 \\
29 & Bootes~II  & 		& 209.5141 &  +12.8553 &  353.7311 &   +68.8649 & 0.026 &  42.0 &  39.8 &     6.6 &   --1.6 &    39.2 & 24 \\
30 & Bootes~I  & 		& 210.0200 &  +14.5135 &  358.1017 &   +69.6366 & 0.015 &  66.0 &  63.5 &    14.4 &   --0.8 &    61.9 & 25 \\
31 & NGC~5694  & 		& 219.9019 & --26.5391 &  331.0560 &   +30.3594 & 0.086 &  35.0 &  29.1 &    17.9 &  --14.6 &    17.7 & 26 \\
32 & Mu\~noz~1  & 		& 225.4490 &  +66.9682 &  105.4414 &   +45.4808 & 0.021 &  45.0 &  47.3 &  --16.9 &    30.4 &    32.1 & 27 \\
33 & NGC~5824  & 		& 225.9943 & --33.0685 &  332.5548 &   +22.0702 & 0.148 &  32.1 &  25.6 &    17.9 &  --13.7 &    12.1 & 28 \\
34 & Ursa Minor  & UGC 9749, DDO 199 & 227.24 &  +67.2221 &  104.9818 &   +44.8127 & 0.028 &  76.0 &  78.0 &  --22.4 &    52.1 &    53.6 & 8 \\
35 & Palomar~14  & AvdB & 242.7544 &  +14.9584 &   28.7472 &   +42.1902 & 0.030 &  76.5 &  71.3 &    41.2 &    27.2 &    51.4 & 29 \\
36 & Hercules  & 		& 247.7722 &  +12.7852 &   28.7275 &   +36.8564 & 0.053 & 132.0 & 126.2 &    84.1 &    50.8 &    79.2 & 15 \\
37 & NGC~6229  & 		& 251.7454 &  +47.5276 &   73.6383 &   +40.3064 & 0.020 &  30.5 &  29.9 &   --2.0 &    22.3 &    19.7 & 30 \\
38 & Palomar~15  & UGC 10642 & 254.9626 &   --0.5390 &   18.8485 &   +24.3370 & 0.338 &  45.1 &  38.0 &    30.4 &    13.3 &    18.6 & 31 \\
39 & Draco & UGC 10822, DDO 208 & 260.0684 &  +57.9185 &   86.3710 &   +34.7128 & 0.024 &  76.0 &  76.0 &   --4.5 &    62.4 &    43.3 & 8 \\ 
40 & NGC~7006 & 		& 315.3722 &  +16.1871 &   63.7691 &  --19.4068 & 0.071 &  41.2 &  38.4 &     8.7 &    34.9 &  --13.7 & 32 \\
41 & Segue~3 & 		& 320.3795 &  +19.1178 &   69.3998 &  --21.2723 & 0.084 &  27.0 &  25.5 &     0.3 &    23.6 &   --9.8 & 33 \\
42 & Pisces~II & 		& 344.6345 &   +5.9526 &   79.2125 &  --47.1089 & 0.056 & 182.0 & 181.1 &    14.7 &   121.7 & --133.3 & 33 \\
43 & Palomar~13 & UGCA 435 & 346.6858 &  +12.7712 &   87.1038 &  --42.7007 & 0.098 &  26.0 &  27.1 &   --7.5 &    19.1 &  --17.6 & 7,8 \\
44 & NGC~7492 & 		& 347.1102 & --15.6108 &   53.3865 &  --63.4764 & 0.031 &  26.3 &  25.4 &   --1.5 &     9.4 &  --23.5 & 34 \\
\enddata
%\tablenotetext{$\dagger$}{~References. -- 
\tablecomments{~References. --
(1) \citet{shapley38a};
(2) \citet{whiting02};
(3) \citet{belokurov09a}; 
(4) \citet{shapley38b};
(5) \citet{lauberts76};
(6) \citet{cesarsky77};
(7) \cite{abell55}; 
(8) \citet{wilson55}; 
(9) \citet{cannon77}; 
(10) Herschel (1788);
(11) \citet{koposov07b}; 
(12) \citet{zucker06b};
(13) \citet{weinberger95}; 
(14) \citet{irwin07a}; 
(15) \citet{belokurov07a};
(16) \citet{harrington50}; 
(17) \citet{irwin90}; 
(18) \citet{willman05b}; 
(19) \citet{willman05a}; 
(20) \citet{belokurov08a}; 
(21) \citet{sakamoto06a};
(22) \citet{zucker06a};
(23) \citet{madore82}; 
(24) \citet{walsh07a}; 
(25) \citet{belokurov06a}; 
(26) Herschel (1784); 
(27) \citet{munoz12a}; 
(28) Dunlop (1826); 
(29) \citet{arp60}; 
(30) Herschel (1787); 
(31) \citet{zwicky59}; 
(32) Herschel (1784); 
(33) \citet{belokurov10a}; 
(34) Herschel (1786). 
}
\end{deluxetable*}

\clearpage

\begin{deluxetable*}{lrcrrrrcrrrrrl}
\tablewidth{\textwidth}
\tabletypesize{\footnotesize}
\tablecaption{Basic Parameters for Outer Halo Objects: Secondary Sample\label{tab:samp2}}
%\tabletypesize{\scriptsize}
%\tablewidth{0pt}
\tablehead{
\colhead{No.} &
\colhead{Name} &
\colhead{Other} &
\colhead{$\alpha$(2000)} &
\colhead{$\delta$(2000)} &
\colhead{$l$} &
\colhead{$b$} &
\colhead{$E(B$-$V)$} &
\colhead{~$R_{\odot}$} &
\colhead{~$R_{\rm GC}$} &
\colhead{~$X$} &
\colhead{~$Y$} &
\colhead{~$Z$} &
\colhead{Reference$^\dagger$}  \\
\colhead{} &
\colhead{} &
\colhead{} &
\colhead{(deg)} &
\colhead{(deg)} &
\colhead{(deg)} &
\colhead{(deg)} &
\colhead{(mag)} &
\colhead{(kpc)} &
\colhead{(kpc)} &
\colhead{(kpc)} &
\colhead{(kpc)} &
\colhead{(kpc)} &
\colhead{} 
}
\startdata
~1 &  Triangulum~II       & Laevens~2 &  33.3252 &    +36.1702 &   140.9044 &   ---23.8281 &    0.068 &     30.0 &     36.5 &    --29.8 &     17.3 &    --12.1  &  1\\
~2 &  Eridanus~3      & &   35.6952 &   --52.2838 &   274.9547 &   ---59.5966 &    0.022 &     87.0 &     87.0 &    --4.6 &    --43.9 &    --75.0  & 2,3 \\
~3 &  Horologium~I & &   43.8812 &   --54.1160 &   271.3843 &   --54.7350 &    0.013 &     79.0 &     79.3 &     --7.3 &    --45.6 &    --64.5  & 2,3 \\
~4 &  Horologium~II & &   49.1076 &   --50.0486 &   262.5314 &   --54.1391 &    0.018 &     78.0 &     79.1 &     --14.3 &    --45.3 &    --63.2  & 4 \\
~5 &  Reticulum~II    & &   53.9203 &   --54.0513 &   266.3007 &   --49.7376 &    0.016 &     30.0 &     31.5 &     --9.7 &    --19.4 &    --22.9  &  2,3 \\
~6 &  Eridanus~II    & &   56.0925 &   --43.5329 &   249.7802 &   --51.6431 &    0.008 &    380.0 &    381.9 &    --89.7 &   --221.4 &   --298.0  & 2,3 \\
~7 &  Pictoris~I        & &   70.9490 &   --50.2854 &   257.3020 &   --40.6438 &    0.011 &    114.0 &    115.7 &    --27.4 &    --84.4 &    --74.3  & 2,3 \\
~8 &  Crater~~~      & Laevens 1 &  174.0668 &   --10.8772 &   274.8070 &   +47.8473 &    0.023 &    170.0 &    169.9 &   1.2 &   --113.7 &    126.0  & 5,6 \\
     &                            & PSO J174.0675-10.8774 & & & & & & & & & & & \\
~9 &  Hydra~II          & &  185.4251 &   --31.9860 &   295.6171 &   +30.4630 &    0.052 &    134.0 &    131.1 &     41.6 &   --104.1 &     67.9  & 7 \\
10 &  Indus~1         &  Kim 2 &  317.2020 &   --51.1671 &   347.1550 &   -42.0692 &    0.026 &    100.0 &     94.0 &     63.9 &    --16.4 &    --67.0  & 8,2,3 \\
11 &  Balbinot~1    & &  332.6791 &    +14.9403 &    75.1723 &   --32.6441 &    0.051 &     31.9 &     31.2 &      --1.7 &     26.0 &    --17.2  & 9 \\
12 &  Kim~1            & &  332.9214 &     +7.0271 &    68.5158 &   --38.4240 &    0.070 &     19.8 &     19.2 &     --2.8 &     14.4 &    --12.3  & 10 \\
13 &  Grus~I          & &  344.1798 &   --50.1800 &   338.6512 &   --58.2367 &    0.007 &    120.0 &    116.1 &     50.4&    --22.9 &   --102.0  & 3 \\
14 &  Phoenix~2    & &  354.9960 &   --54.4115 &   323.6820 &   --59.7438 &    0.012 &     83.0 &     79.9 &     25.2 &    --24.7 &    --71.7  & 2,3 \\
\enddata
%\tablenotetext{$\dagger$}{~References. -- 
\tablecomments{~References. --
(1) \citet{laevens15a}; (2) \citet{bechtol15a}; (3) \citet{koposov15a}; (4) \citet{kimjerjen15b}; 
(5) \citet{laevens14}; (6) \citet{belokurov14}; (7) \citet{martin15}; (8) \citet{kim15a}; (9) \citet{balbinot13}; 
(10) \citet{kimjerjen15a}; (11) \citet{kim15b}.}
\end{deluxetable*}

%\clearpage
%\end{landscape}
 
\begin{deluxetable*}{cccccccccc}
\tablecaption{Observing Run Information
\label{tab:runs}}
\tabletypesize{\footnotesize}
\tablewidth{0pt}
\tablehead{
\colhead{Telescope} &
\colhead{Instrument} &
\colhead{PI} &
\colhead{N$_{\rm CCD}$} &
\colhead{Field} &
\colhead{Scale} &
\colhead{FWHM} &
\colhead{Filters} &
\colhead{ID \#} &
\colhead{Date} \\
\colhead{} &
\colhead{} &
\colhead{} &
\colhead{} &
\colhead{(deg)} &
\colhead{(\arcsec/pixel)} &
\colhead{(\arcsec)} &
\colhead{} &
\colhead{} &
\colhead{} 
}
\startdata
CFHT & MegaCam  & C\^ot\'e & 36 & 0.96$\times$0.94 & 0.187 & 0.7--0.9 & $g,r$ & 09AC07 & 2009-A \\
& & & & & & & & 09BC02 & 2009-B \\
& & & & & & & & 10AC06 & 2010-A \\
& & & & & & & & & \\
Clay   & MegaCam  & Geha    & 36 & 0.40$\times$0.40 & 0.160 & 0.7--1.1 & $g,r$ & 2010B-0472 & 2010-B \\
& & Simon & & & & & & \nodata & 2010-B \\
& & Simon & & & & & & \nodata & 2011-A \\
\enddata
\end{deluxetable*}

\clearpage

\begin{deluxetable*}{llrccccccrr}
\tablecaption{Summary of Observations for Primary Sample Objects
\label{tab:obslog}}
\tabletypesize{\footnotesize}
\tablewidth{0pt}
\tablehead{
\colhead{No.} &
\colhead{Name} &
\colhead{Telescope} &
\colhead{Mosaic} &
\colhead{Area} &
\colhead{Astrometry} &
\colhead{Photometry} &
\colhead{$\langle{X}\rangle_{g}$} &
\colhead{$\langle{X}\rangle_{r}$} &
\colhead{$T_g$~} &
\colhead{$T_r$~} \\
\colhead{} &
\colhead{} &
\colhead{} &
\colhead{} &
\colhead{(deg$^2$)} &
\colhead{} &
\colhead{} &
\colhead{} &
\colhead{} &
\colhead{(sec)} &
\colhead{(sec)} 
}
\startdata
~1 &  Sculptor     	& Clay		& $4\times3$ 	& 1.13 & GAIA-DR1 & secondary &  $1.05$ & $1.05$ & $5\times 90$ & $5\times 180$\\
~2 &  Whiting 1     & CFHT	& $1\times1$ 	& 0.95 & GAIA-DR1 & secondary &  $1.31$ & $1.20$ & $6\times 300$ & $6\times 300$\\
~3 &  Segue~2      & CFHT	& $1\times1$   & 1.02 & SDSS-DR7 & SDSS        &  $1.19$ & $1.35$ & $6\times 350$ & $6\times 350$\\
~4 &  Fornax     	& Clay		& $2\times2$ 	& 0.62 & GAIA-DR1 & secondary  &  $1.01$ & $1.01$ & $5\times 90$ & $5\times 180$\\
~5 &  AM~1     	& Clay		& $1\times1$   & 0.21 & GAIA-DR1 & SDSS &  $1.07$ & $1.07$ & $5\times 90$ & $5\times 180$\\
~6 &  Eridanus     	& CFHT	& $1\times1$ 	& 0.94 & USNO-B1 & secondary &  $1.33$ & $1.33$ & $6\times 270$ & $6\times 270$\\
~7 &  Palomar 2   	& CFHT	& $1\times1$ 	& 0.95 & USNO-B1 & secondary &  $1.14$ & $1.31$ & $6\times 450$ & $6\times 450$  \\
~8 &  Carina     	& Clay		& $4\times4$   & 2.16 & GAIA-DR1 & secondary &  $1.15$ & $1.15$ & $5\times 90$ &$5\times 180$\\
~9 &  NGC 2419  	& CFHT	& $1\times1$ 	& 1.02 & SDSS-DR7 & SDSS &  $1.40$ & $1.17$ & $6\times 450$ & $6\times 450$\\
10 &  Koposov 2    & CFHT	& $1\times1$ 	& 0.94 & SDSS-DR7 & SDSS &  $1.28$ & $1.26$ & $6\times 500$ & $6\times 500$\\
11 &  UMa II       	& CFHT	& $2\times2$ 	& 2.77 & GAIA-DR1 & SDSS &  $1.39$ & $1.39$ & $11\times 270$ & $11\times 468$\\
12 &  Pyxis     		& Clay		& $1\times1$  	& 0.20 & GAIA-DR1 & SDSS &  $1.16$  & $1.14$ & $5\times 180$ & $5\times 180$\\
13 &  Leo~T     	 & Clay		& $1\times1$   & 0.21 & GAIA-DR1 & SDSS &  $1.45$  & $1.47$ & $5\times 180$ & $5\times 180$\\
14 &  Palomar 3   	& CFHT	& $1\times1$ 	& 0.98 & GAIA-DR1 & SDSS &  $1.27$ & $1.42$ & $6\times 270$ & $6\times 270$ \\
     &                      	& Clay		& $1\times1$	& 0.21 & GAIA-DR1 & SDSS &  $1.18$ & $1.17$ &$5\times 90$ & $5\times 180$ \\
15 &  Segue~1     	& CFHT	& $1\times1$ 	& 0.95 & GAIA-DR1 & SDSS &  $1.19$ & $1.37$& $6\times 60$ & $6\times 60$\\
16 &  Leo~I 		& CFHT	& $1\times1$ 	& 1.04 & GAIA-DR1 & SDSS &  $1.25$ & $1.10$  & $11\times 300$ & $11\times 360$\\
17 &  Sextans 		& CFHT	& $2\times2$ 	& 3.30 & GAIA-DR1 & SDSS &  $1.10$ & $1.16$ & $6\times 310$ & $6\times 310$\\ 
18 &  UMa~I 		& CFHT	& $2\times1$ 	& 0.98 & GAIA-DR1 & SDSS &  $1.24$ & $1.21$ & $6\times 600$ & $6\times 600$\\
19 &  Willman~1 	& CFHT	& $1\times1$ 	& 0.94 & GAIA-DR1 & SDSS &  $1.40$ & $1.37$ & $6\times 180$ & $6\times 180$ \\
20 &  Leo~II     	& Clay		& $2\times2$   & 0.15 & GAIA-DR1 & SDSS &  $1.65$ & $1.65$ & $5\times 90$ & $5\times 180$\\
21 &  Palomar 4   	& CFHT	& $1\times1$ 	& 0.96 & GAIA-DR1 & SDSS &  $1.24$ & $1.31$ & $6\times 440$ & $6\times 440$ \\
22 &  Leo~V     	& Clay		& $1\times1$   & 0.21 & GAIA-DR1 & SDSS &  $1.20$ & $1.18$ & $5\times 90$ & $5\times 180$\\
23 &  Leo~IV  	 	& Clay		& $1\times1$ 	& 0.21 & GAIA-DR1 & SDSS &  $1.15$ & $1.14$ & $5\times 90$ & $5\times 180$\\
24 &  Koposov~1   & Clay		& $1\times1$   & 0.21 & GAIA-DR1 & SDSS &  $1.33$  & $1.33$ & $5\times 90$ & $5\times 180$\\
25 &  ComBer     	& CFHT	& $2\times2$ 	& 3.51 & GAIA-DR1 & SDSS &  $1.10$  & $1.10$ & $11\times 270$ & $11\times 468$\\
26 &  CVn~II		& CFHT	& $1\times1$ 	& 1.01 & GAIA-DR1 & SDSS &  $1.44$ & $1.38$ & $6\times 380$ & $6\times 380$ \\
27 &  CVn~I		& CFHT	& $1\times1$ 	& 0.96 & GAIA-DR1 & SDSS &  $1.65$ & $1.56$ & $9\times 300$ & $9\times 450$ \\
28 &  AM~4     		& Clay		& $1\times1$ 	& 0.21 & GAIA-DR1 & SDSS &  $1.04$ & $1.03$ & $5\times 90$ & $5\times 180$\\
29 &  Bootes~II 	& CFHT	& $1\times1$ 	& 0.97 & GAIA-DR1 & SDSS &  $1.24$ & $1.23$ & $6\times 400$ & $6\times 400$\\
30 &  Bootes~I  	& CFHT	& $1\times2$ 	& 1.59 & GAIA-DR1 & SDSS &  $1.30$ & $1.30$ & $6\times 140$ & $6\times 140$\\
31 &  NGC 5694    & CFHT	& $1\times1$  	& 0.95 & GAIA-DR1 & secondary &  $1.45$ & $1.45$& $6\times 60$ & $6\times 60$\\ 
32 &  Mu\~noz~1 	& CFHT	& $1\times1$ 	& \nodata & GAIA-DR1 & SDSS &  $1.54$ & $1.50$ & $6\times 200$ & $6\times 200$\\
33 &  NGC~5824   & Clay		& $1\times1$   & 0.21 & GAIA-DR1 & SDSS &  $1.08$ & $1.06$ & $5\times 90$ & $5\times 180$\\
34 &  Ursa Minor  	& CFHT	& $2\times2$ 	& 3.21 & GAIA-DR1 & SDSS &  $1.54$ & $1.50$ & $6\times 200$ & $6\times 200$\\
35 &  Palomar~14 & CFHT	& $1\times1$   & 0.95 & GAIA-DR1 & SDSS &  $1.33$ & $1.25$ & $6\times 120$ & $6\times 120$ \\
36 &  Hercules       & CFHT	& $2\times1$ 	& 2.01 & GAIA-DR1 & SDSS &  $1.11$ & $1.14$ & $6\times 400$ & $6\times 400$\\
37 &  NGC~6229   & CFHT	& $1\times1$ 	& 0.91 & GAIA-DR1 & SDSS &  $1.34$ & $1.39$& $6\times 90$ & $6\times 90$  \\
38 &  Palomar~15  & CFHT	& $1\times1$   & 1.03 & GAIA-DR1 & SDSS &  $1.09$ & $1.13$ & $6\times 2250$ & $6\times 225$\\
39 &  Draco 		 & CFHT	& $2\times2$ 	& 3.34 & GAIA-DR1 & SDSS &  $1.48$ & $1.37$ & $6\times 150$ & $6\times 125$\\
40 &  NGC~7006   & CFHT	& $1\times1$ 	& 0.95 & GAIA-DR1 & SDSS &  $1.01$ & $1.01$ & $6\times 240$ & $6\times 240$\\
41 &  Segue~3       & Clay  	& $1\times1$ 	& 0.21 & GAIA-DR1 & SDSS &  1.72 & 1.65 & $5\times 90$ & $5\times 180$ \\
42 &  Pisces~II 	  & Clay   	& $1\times1$   & 0.21 & GAIA-DR1 & SDSS    &  1.23 & 1.26 & $5\times 225$ & $5\times 450$ \\
43 &  Palomar~13   & CFHT	& $1\times1$   & 0.98 & GAIA-DR1 & SDSS   &  $1.05$ & $1.01$ & $6\times 360$ & $6\times 360$\\
44 &  NGC~7492    & CFHT	& $1\times1$ 	& 0.94 & GAIA-DR1 & secondary &  $1.23$ & $1.23$ & $6\times 120$ & $6\times 120$\\
&                      	 & Clay		& $1\times1$	& 0.21 & GAIA-DR1  & secondary 	&  $1.07$ & $1.09$ 	& $5\times 90$ & $5\times 180$ \\
\enddata 
\end{deluxetable*}

\clearpage

\begin{deluxetable*}{ccccccccr}
\scriptsize
\tablewidth{0pt}
\tablecaption{Photrometry of Individual Objects}
\tablehead{
\colhead{Star ID} &
\colhead{RA(deg, $2000.0$)} &
\colhead{DEC(deg, $2000.0$)} &
\colhead{$g$} &
\colhead{$\sigma_{g}$} &
\colhead{$r$} &
\colhead{$\sigma_{r}$} &
\colhead{$chi$} &
\colhead{$sharp$}
}
\startdata
scl$-1$ &  $15.067787$	& $-33.383965$ 	& $20.9255$	&	 $0.0023$   & $20.4028$  	&  $0.0018$ & $0.5097$  	&  $0.0035$ \\
scl$-2$ &  $15.102280$	& $-33.383710$ 	& $23.2864$	&	 $0.0119$  & $23.0176$  	&  $0.0094$ & $0.4387$  	&  $-0.0646$ \\
scl$-3$ &  $15.101350$	& $-33.382940$ 	& $23.3046$	&	 $0.0121$  & $22.9910$  	&  $0.0092$  & $0.4393$  	&  $0.0703$\\ 
scl$-4$ &  $15.098076$ & $-33.382812$ 	& $22.9103$ 	&	$0.0094$	&  $22.4541$ 	&  $0.0068$  & $0.4669$  &  $0.6238$\\
\enddata
%\smallskip
\label{t:photometry_all}
\tablecomments{Only a portion of this table is shown here to demonstrate its form and content. A machine-readable version of the full table containing the
photometry for all primary objects is available.
Objects from different satellites can be separated by their ID name.} 
\end{deluxetable*}

%{\scriptsize
%{Only a portion of this table is shown here to demonstrate its form and content. A machine-readable version of the full table containing the
%photometry for all primary objects is available.
%Objects from different satellites can be separated by their ID name.} 
%}

\clearpage
%\vskip -1.2cm

\begin{deluxetable*}{llrccccccrr}
\tablecaption{Summary of Observations for Secondary Sample Objects
\label{tab:obslog2}}
\tabletypesize{\footnotesize}
\tablewidth{0pt}
\tablehead{
\colhead{No.} &
\colhead{Name} &
\colhead{Telescope} &
\colhead{Mosaic} &
\colhead{Area} &
\colhead{Astrometry} &
\colhead{Photometry} &
\colhead{$\langle{X}\rangle_{g}$} &
\colhead{$\langle{X}\rangle_{r}$} &
\colhead{$T_g$~} &
\colhead{$T_r$~} \\
\colhead{} &
\colhead{} &
\colhead{} &
\colhead{} &
\colhead{(deg$^2$)} &
\colhead{} &
\colhead{} &
\colhead{} &
\colhead{} &
\colhead{(sec)} &
\colhead{(sec)} 
}
\startdata
~1 &  Triangulum~II     	& LBT	& $1\times1$ 	& $23\arcmin \times25\arcmin$ & PS1 & PS1 &  \nodata & \nodata & $6\times 200$ & $6\times 200$\\
~2 &  Eridanus~3    		& Blanco	& $1\times1$ 	& $\sim3$ & DESDM & secondary &  $1.09$ & $1.10$ & $1\times90$ & $1\times90$\\
~3 &  Horologium~I      	& Blanco	& $1\times1$  	& $\sim3$ & DESDM & secondary        &  $1.19$ & $1.40$ & $1\times90$ & $1\times90$\\
~4 &  Horologium~II     	& Blanco	& $1\times1$ 	& $\sim3$ & DESDM & APASS DR8  &  \nodata & \nodata & $1\times 90$ & $1\times 90$\\
~5 &  Reticulum~II    		& Blanco	& $1\times1$   	& $\sim3$ & DESDM & secondary &  $1.38$ & $1.39$ & $1\times90$ & $1\times90$\\
~6 &  Eridanus~II     		& Blanco	& $1\times1$ 	& $\sim3$ & DESDM & secondary &  $1.31$ & $1.32$ & $1\times90$ & $1\times90$\\
~7 &  Pictoris~I   		& Blanco	& $1\times1$ 	& $\sim3$ & DESDM & secondary &  $1.09$ & $1.10$ & $1\times90$ & $1\times90$  \\
~8 &  Crater    			& PS1 	& \nodata   	& \nodata & PS1 & PS1 &  \nodata & \nodata & $5\times 90$ &$5\times 180$\\
~9 &  Hydra~II  			& Blanco	& $1\times1$ 	& $\sim3$ & DESDM & \nodata &  \nodata & \nodata & $3\times 267$ & $3\times 267$\\
10 &  Indus~1    		& Blanco	& $1\times1$ 	& $\sim3$ & DESDM & secondary &  $1.20$ & $1.07$ & $1\times90$ & $1\times90$\\
11 &  Balbinot~1       		& CFHT	& $1\times1$ 	& $\sim1$ & \nodata & SDSS &  \nodata & \nodata & $6\times 467$ & $6\times 633$\\
12 &  Kim~1     			& Blanco	& $1\times1$  	& $\sim3$ & DESDM & SDSS &  \nodata  & \nodata & $8\times 210$ & $5\times 210$\\
13 &  Grus~I     		& Blanco	& $1\times1$ 	& $\sim3$ & DESDM & secondary &  $1.10$ & $1.12$& $1\times90$ & $1\times90$\\
14 &  Phoenix~2 		& Blanco	& $1\times1$ 	& $\sim3$ & DESDM & secondary &  $1.17$ & $1.38$  & $1\times90$ & $1\times90$\\
\enddata 
\end{deluxetable*}

%%%%%%%%%%%%%%%%%%%%%%%%%%%%%%%%%%%%%%%%%%%%%%%%%%%%%%%%%%%%%%%%%%%%%%
%   FIGURES
%%%%%%%%%%%%%%%%%%%%%%%%%%%%%%%%%%%%%%%%%%%%%%%%%%%%%%%%%%%%%%%%%%%%%%

\end{document}